\documentclass[journal,twoside,web]{ieeecolor}
\usepackage{generic}
\usepackage{cite}
\usepackage{amsmath,amssymb,amsfonts}
\usepackage{graphicx}
\usepackage{algorithm}
\usepackage{hyperref}
\hypersetup{hidelinks=true}
\usepackage{textcomp}
\usepackage{orcidlink}
\usepackage{amssymb}
\usepackage{amsmath}
\usepackage{ wasysym }

\newcommand{\defdone}[0]{\hfill \lhd} 

\newcommand{\qed}[0]{\hfill \square}

\usepackage{cases}
\usepackage{empheq}
\usepackage{mathbbol}
\usepackage{bm}
\usepackage{bbm}
\usepackage{mathtools}
\usepackage{mathtools}
\usepackage{accents}

\usepackage{siunitx}
\usepackage{subcaption}  
\usepackage{cite}

\usepackage{accents}
\usepackage{algpseudocode}

\graphicspath{./Figures/}

\newcommand{\reals}{\mathbb{R}}

\newcommand{\naturals}{\mathbb{N}}
\newcommand{\Ball}{\mathcal{B}}

\newcommand{\TS}{S}
\newcommand{\ConStateSet}{X}
\newcommand{\ConState}{x}
\newcommand{\ConFlow}{f}

\newcommand{\DimSys}{n}

\newcommand{\AbsTS}{\hat{\TS}}
\newcommand{\AbsStateSet}{\hat{\ConStateSet}}
\newcommand{\AbsState}{\hat{\ConState}}
\newcommand{\AbsFlow}{\hat{\ConFlow}}
\newcommand{\CardAbs}{k}
\newcommand{\Ancestors}{\mathcal{A}}
\newcommand{\Descendants}{\mathcal{D}}

\newcommand{\Quantizer}{\kappa}
\newcommand{\Mesh}{\mathcal{M}}
\newcommand{\MeshCell}{C}
\newcommand{\dist}{d}
\newcommand{\RelCell}{L}
\newcommand{\Params}{\theta}

\newcommand{\Rel}{R}

\newcommand{\PropCoef}{u}
\newcommand{\AdditCoef}{v}

\newcommand{\Data}{D}
\newcommand{\SubProb}{\Params^*}
\newcommand{\SolSpace}{\mathcal{W}}
\newcommand{\Refine}{\text{Refine}}

\newtheorem{definition}{Definition}
\newtheorem{proposition}{Proposition}
\newtheorem{corollary}{Corollary}
\newtheorem{theorem}{Theorem}
\newtheorem{assumption}{Assumption}
\newtheorem{example}{Example}
\newtheorem{lemma}{Lemma}
\newtheorem{remark}{Remark}
\newtheorem{problemstatement}{Problem Statement}
\newtheorem{problemrelaxation}{Relaxed Problem}

\usepackage{url}

\def\BibTeX{{\rm B\kern-.05em{\sc i\kern-.025em b}\kern-.08em
    T\kern-.1667em\lower.7ex\hbox{E}\kern-.125emX}}
\begin{document}
\title{Existence and Synthesis of Multi-Resolution Approximate Bisimulations for Continuous-State Dynamical Systems}
\author{Rudi Coppola$^{\orcidlink{0000-0002-1876-6827}}$, \IEEEmembership{Student, IEEE}, Yannik Schnitzer$^{\orcidlink{0000-0001-7406-3440}}$, Mirco Giacobbe$^{\orcidlink{0000-0001-8180-0904}}$, Alessandro Abate$^{\orcidlink{0000-0002-5627-9093}}$, \IEEEmembership{Fellow, IEEE}, and Manuel Mazo Jr.$^{\orcidlink{0000-0002-5638-5283}}$, \IEEEmembership{Senior Member, IEEE}
\thanks{Rudi Coppola and Manuel Mazo Jr. are with the Delft University of Technology, Delft, Netherlands (e-mail: r.coppola@tudelft.nl, m.mazo@tudelft.nl). }
\thanks{Yannik Schnitzer and Alessandro Abate are with the University of Oxford, Oxford, United Kingdom (e-mail: yannik.schnitzer@cs.ox.ac.uk,alessandro.abate@cs.ox.ac.uk).}
\thanks{Mirco Giacobbe is with 
the University of Birmingham, Birmingham, United Kingdom (e-mail: m.giacobbe@bham.ac.uk).}}

\maketitle
\begin{abstract}

We present a fully automatic framework for synthesising compact, finite-state deterministic abstractions of deterministic, continuous-state autonomous systems under locally specified resolution requirements.

Our approach builds on multi-resolution approximate bisimulations, a generalisation of classical $\epsilon$-approximate bisimulations, that support state-dependent error bounds and subsumes both variable- and uniform-resolution relations. We show that some systems admit multi-resolution bisimulations but no $\epsilon$-approximate bisimulation.

We prove the existence of multi-resolution approximately bisimilar abstractions for all incrementally uniformly bounded ($\delta$-UB) systems, thereby broadening the applicability of symbolic verification to a larger class of dynamics; as a trivial special case, this result also covers incrementally globally asymptotically stable ($\delta$-GAS) systems.

The Multi-resolution Abstraction Synthesis Problem (MRASP) is solved via a scalable Counterexample-Guided Inductive Synthesis (CEGIS) loop, combining mesh refinement with counterexample-driven refinement. This ensures soundness for all $\delta$-UB systems, and ensures termination in certain special cases.

Experiments on linear and nonlinear benchmarks, including non-$\delta$-GAS and non-differentiable cases, demonstrate that our algorithm yields abstractions up to 50\% smaller than Lyapunov-based grids while enforcing tighter, location-dependent error guarantees.

\end{abstract}


%
\section{Introduction} Formal verification and synthesis of control systems have become essential in ensuring safety and reliability in safety-critical domains such as robotics, autonomous driving, and cyber-physical systems. The key challenge is that continuous-state dynamical systems are often too complex for direct analysis, requiring the construction of finite-state abstractions that preserve essential properties of the original system. Symbolic methods based on bisimulation relations have been highly successful in bridging this gap between continuous dynamics and discrete verification frameworks. Essential to these developments is the notion of approximate relations that relax the strict requirements of exact bisimulation~\cite{DBLP:journals/automatica/Pappas03,DBLP:conf/hybrid/Schaft04,DBLP:journals/tac/Schaft04}. Early work introduced $\epsilon$-approximate bisimulation and bisimulation functions, where system behaviours are matched within a uniform error margin \cite{tabuada2009verification, girard2011approximate, girard2005approximate}. 
\newline
Classical approaches rely on uniform-resolution abstractions, typically based on gridding the state space. While such methods provide rigorous guarantees, they often suffer from scalability issues due to the curse of dimensionality. Moreover, many of these techniques require the existence of certificates, such as Lyapunov-like functions \cite{girard2009approximately, zamani2011symbolic}, which can be difficult to obtain in practice. This motivates the search for abstraction techniques that are both more flexible and applicable to a broader class of systems. To improve scalability, \cite{camara2011safety,camara2011synthesis,girard2015safety, hsu2018multi} proposed abstractions using a set of grids of different coarseness, leading to more compact abstractions. In parallel, the notion of multi-resolution approximate bisimulation has been proposed to exploit heterogeneous error bounds across the state space, allowing for finer approximations in regions of interest while maintaining coarser approximations elsewhere \cite{tazaki2010approximately,tazaki2011discrete}. \newline
To alleviate the computational complexity of abstracting systems when no certificates are known, data-driven and learning-based abstraction methods such as \cite{abate2024bisimulation,abate2022neural,DBLP:conf/cav/AbateGMS25,nadali2024transfer}, rely on a data set of samples of the dynamics, such as transitions for discrete-time systems, to train parametrised functions which elicit a formal relation between the original system and a candidate abstraction. In the following, we provide a detailed comparison with the relevant existing literature.

\textbf{Related work}: Mitigating the curse of dimensionality is a central challenge in abstractions for dynamical systems and has received significant attention over the past decade. Instead of a uniform grid, \cite{camara2011safety,camara2011synthesis} introduce multi-scale abstractions by combining grids of different coarseness, extending \cite{girard2009approximately} to $\epsilon$-approximate bisimulations for $\delta$-GAS switched systems. This allows for generating abstraction with smaller state sets, compared to the original approach based on a uniform grid, that are suitable for safety and reachability synthesis.. Similarly, \cite{hsu2018multi} exploit multi-scale grids for systems with known growth bounds, using feedback-refinement relations (FRR) to handle $\omega$-regular specifications. In \cite{weber2016optimized}, a functional predicts abstraction costs for FRR and guides the selection of hyper-intervals to minimise transitions. FRR also underpins \cite{calbert2024smart}, where sparse, goal-specific abstractions are built via backward search and ellipsoidal coverings of $L$-smooth systems with affine controllers, yielding compact abstractions. Several of these methods are integrated in the tool \cite{calbert2024dionysos}.

In contrast, we employ multi-resolution approximate bisimulations, first proposed in \cite{tazaki2010approximately}, where a resolution function defines heterogeneous precision. The authors use local linearisation (under differentiability), a candidate abstraction, and a norm-based template to cast the existence of such an abstraction as a linear program, refined by a greedy heuristic if infeasible. Their method is efficient for linear systems, though local linearisations can be costly \cite{tazaki2011discrete}. Our approach avoids linearisations, relying instead on sampled transitions, thus applying to a broader class of systems.

In data-driven settings, \cite{nadali2024transfer} introduces neural simulation relations: given a discretised abstraction, relations and controllers are parameterised as neural networks, trained on system transitions, and verified via Lipschitz continuity for $\epsilon$-approximate simulation. \cite{abate2022neural} abstract continuous-time deterministic systems by sampling the vector field to train a neural network, producing hybrid automata with linear disturbed dynamics verifiable using SMT. More recently, \cite{abate2024bisimulation, DBLP:conf/cav/AbateGMS25} abstract discrete-time and discrete-state systems into stutter bisimilar models through SMT queries over transition data. Analogously, we synthesise abstractions and multi-resolution approximate bisimulations as SMT queries, solved via CEGIS.


\textbf{Main Contributions}: This work provides the first existential results for multi-resolution approximate bisimulation relations. Since this notion strictly generalises $\epsilon$-approximate bisimulations, any $\delta$-GAS system admits a multi-resolution bisimilar abstraction. More significantly, we show that $\delta$-UB systems, a strictly larger class, also admit such abstractions.

We then formalise the Multi-resolution Abstraction Synthesis Problem (MRASP), cast it as a satisfiability problem in first-order logic, and solve it efficiently through a fully automatic Counterexample-Guided Inductive Synthesis (CEGIS) loop, which iteratively refines both the abstraction and its relation until validity is achieved. Unlike existing methods, our approach requires neither $\delta$-GAS Lyapunov functions nor local linearisation, making it applicable to non-$\delta$-GAS and even non-differentiable systems. 

The synthesis algorithm is parallelisable, leveraging the abstraction’s graph structure to distribute refinement and verification. Empirical results show that the produced abstractions are substantially more compact than grid- or Lyapunov-based ones, while location-dependent error bounds yield tighter guarantees than uniform resolutions. Benchmarks on linear, nonlinear, and non-differentiable systems confirm scalability and soundness, providing a fully automated tool for constructing abstractions under local resolution requirements.

\textbf{Organisation:} We begin in Section \ref{sec:multi-res-bisim}, by defining and motivating our interest in \textbf{multi-resolution approximate bisimulations}. In Section \ref{sec:compact-abstractions}, we formally state the problem of generating \textbf{compact abstractions} that efficiently approximate a system at a specified resolution. To address computational challenges, we relax this problem in Section \ref{sec:cegis-multi-res}, This section introduces our primary algorithmic contribution: a solution based on \textbf{CEGIS}. We then detail two different implementations of the CEGIS loop: Section \ref{sec:low-level-loop-one-shot} describes a one-shot approach and Section \ref{sec:low-level-loop-parall} presents a parallelised version, highlighting its computational advantages. Section \ref{sec:main-result} contains our main theoretical result, establishing that $\boldsymbol{\delta}$\textbf{-UB systems} are guaranteed to admit multi-resolution approximate bisimulations. Finally, in Section \ref{sec:num-examples}, we validate our methodology with several numerical examples.

\section{Preliminaries}\label{sec:prelims}

For a relation $\Rel\subseteq A\times B$ we denote by  
$\Rel(b)\doteq\{a\in A : (a,b)\in\Rel\}$ the projection of $\Rel$ at $b$ on $A$. Given a metric $\dist$ on $\ConStateSet$, we denote an open ball of radius $r$ centred at $\ConState\in\ConStateSet$ by $\Ball_{r}(\ConState)\doteq\{\ConState'\in\ConStateSet : \dist(\ConState,\ConState')<r\}$. For $\ConState \in \reals^\DimSys$, $\ConState[i]$ denotes the $i$-th component, and $\ConState[i:j]$ the subvector consisting of components $i$ through $j$, inclusive. A function $\alpha:\reals_{\ge0}\to\reals_{\ge0}$ is of class $\mathcal{K}$ if it is continuous, strictly increasing, and $\alpha(0)=0$.
A function $\beta:\reals_{\ge0}\times\reals_{\ge0}\to\reals_{\ge0}$ is of class $\mathcal{KL}$ if, for each fixed $t\ge0$, $\beta(\cdot,t)\in\mathcal{K}$, and for each fixed $s\ge0$, $\beta(s,t)\to0$ as $t\to\infty$.

\begin{definition}
    A (simple discrete-time continuous-state deterministic) transition system (TS) $\TS$ is a tuple $(\ConStateSet,\ConStateSet_0,\ConFlow)$ where $\ConStateSet_0\subseteq\ConStateSet\subset\reals^n$ define the (initial) state set and $\ConFlow:\ConStateSet\rightarrow\ConStateSet$ is the transition function. $\defdone$
\end{definition}

For $i\geq1$, we denote by $\ConFlow^{i}$ the $i$-fold composition of $\ConFlow$, i.e. $\ConFlow^{1}=\ConFlow$, $\ConFlow^{2} = \ConFlow\circ\ConFlow$, and so on.

\begin{definition}
    A set $T$ is called \emph{transient} for $\TS$ if there exists $\tau\in\naturals$ called \emph{transient time} such that for all $\ConState\in T$  there exists $q\leq \tau$ such that $\ConFlow^{q}(\ConState)\notin T$. $\defdone$
\end{definition}

\begin{definition}
    Let $\AbsStateSet$ be a collection of $\CardAbs\in\naturals$ elements in $\ConStateSet$. A \emph{quantiser} of $\ConStateSet$ is a function $\Quantizer:\ConStateSet\rightarrow\AbsStateSet$ mapping every element of $\ConStateSet$ to one of the elements of $\AbsStateSet$. $\defdone$
\end{definition}

A quantiser can be equivalently represented by a \emph{mesh}.

\begin{definition}\label{def:quant-to-mesh}
    The mesh of a quantiser $\kappa$ is defined as $\Mesh\doteq\{(\AbsState_1,\MeshCell_1),\ldots(\AbsState_\CardAbs,\MeshCell_\CardAbs)\}$, where $\MeshCell_i$ is the $i$-th cell of the mesh, and $\{\MeshCell_i\doteq\{\ConState\in\ConStateSet : \AbsState_i = \Quantizer(\ConState)\}\}_{i=1}^\CardAbs$ partitions $\ConStateSet$. $\defdone$
\end{definition}

We aim to represent $\TS$ by an \emph{abstraction} $\AbsTS$, a simpler finite-state model whose state set and transition function can be related to the ones of $S$. When such a relation can be determined and quantified, it is possible to establish connections between the behaviours of $\TS$ and those of $\AbsTS$. We restrict our attention to deterministic abstractions.

\begin{definition}\label{def:det-abs}
    A deterministic abstraction of $\TS$ is given by the transition system $\AbsTS\doteq(\AbsStateSet,\AbsStateSet_0,\AbsFlow)$ where $\AbsStateSet_0\doteq\{\AbsState \in\AbsStateSet: \exists \ConState\in\ConStateSet_0 \ . \ \AbsState=\Quantizer(\ConState)\}$ and $\AbsFlow:\AbsStateSet\rightarrow\AbsStateSet:\AbsState\mapsto\Quantizer(f(\AbsState))$. $\defdone$
\end{definition}

\section{Multi-resolution Bisimulation}\label{sec:multi-res-bisim}

\emph{Exact} bisimulation relations between systems require their behaviours to be identical~\cite{DBLP:journals/tac/Schaft04}.
In the context of abstracting continuous-state dynamical systems, this notion is restrictive. Instead, when there is a metric on the behaviours, \emph{approximate} relations have proven to be successful, as they only require the behaviours of the concrete system and its abstraction to not differ more than a prescribed margin of error \cite{tabuada2009verification, girard2009approximately}. Below, we recall the definition of $\epsilon$-approximate bisimulation relation, adapted for deterministic TS.
\begin{definition}[Adapted from \cite{tabuada2009verification}]\label{def:eps-approx}
    A relation $\Rel\subseteq\ConStateSet\times\AbsStateSet$ is an $\epsilon$-approximate bisimulation relation between $\TS$ and $\AbsTS$, written $\TS\simeq_{\Rel}^{\epsilon}\AbsTS$ if
    \begin{enumerate}
        \item for all $\ConState\in\ConStateSet_0$ there exists $\AbsState\in\AbsStateSet_0$ such that $(\ConState,\AbsState)\in\Rel$, and for all $\AbsState\in\AbsStateSet_0$ there exists $\ConState\in\ConStateSet_0$ such that $(\ConState,\AbsState)\in\Rel$,
        \item for all $(\ConState,\AbsState)\in\Rel$ it holds that $(\ConFlow(\ConState),\AbsFlow(\AbsState))\in\Rel$,
        \item for all $(\ConState,\AbsState)\in\Rel$ it holds that $\dist(\ConState,\AbsState)\leq\epsilon$.\footnote{Since we restrict our attention to deterministic concrete systems and deterministic abstractions, condition 2 is stated in a simplified form when compared to the standard $\epsilon$-approximate simulation relation.}$\defdone$
    \end{enumerate} 
\end{definition}
The definition above has been used successfully to prove that Incrementally Globally Asymptotically Stable ($\delta$-GAS) systems admit an abstraction that is $\epsilon$-approximately bisimilar to the system \cite{tabuada2009verification, girard2009approximately}.

This manuscript follows the seminal work of \cite{tazaki2010approximately}, focusing on a generalisation of the notion of $\epsilon$-approximate relations, to construct \emph{compact} abstractions $\AbsTS$. The resulting abstractions must approximate the original system $\TS$ with a sufficiently high resolution provided by the user as a \emph{space-dependent resolution specification}, expressed as a relation between the concrete and abstract states.
Formally, let $\dist$ be a metric on $\ConStateSet$, $\epsilon:\ConStateSet\times\ConStateSet\rightarrow\reals_{>0}$ be a \emph{resolution function}, and $\overline{\Rel}\subseteq \ConStateSet\times\ConStateSet$ be a relation of interest, or \emph{resolution relation}, of the form 
\begin{equation}\label{eq:target-relation}
    \overline{\Rel}\doteq\{(\ConState,\AbsState) \in \ConStateSet\times\ConStateSet: \dist(\ConState,\AbsState) \leq \epsilon(\ConState,\AbsState)\}.
\end{equation}
We define 
sufficient conditions to establish a resolution relation between the TS and its abstraction.

\begin{definition}[Adapted from \cite{tazaki2010approximately}]\label{def:multi-res-simulation}
    A relation $\Rel$ is a multi-resolution approximate simulation relation from $\TS$ to $\AbsTS$ with resolution $\overline{\Rel}$, denoted by $\TS\preceq_{\Rel}^{\overline{\Rel}}\AbsTS$, if:
    \begin{enumerate}
        \item for all $\ConState\in\ConStateSet_0$ there exists $\AbsState\in\AbsStateSet_0$ such that $(\ConState,\AbsState)\in\Rel$,
        \item for all $(\ConState,\AbsState)\in\Rel$ it holds that $(\ConFlow(\ConState),\AbsFlow(\AbsState))\in\Rel$,
        \item $\Rel\subseteq\overline{\Rel}$.$\defdone$
    \end{enumerate}
    \end{definition}
    \begin{definition}\label{def:multi-res-bisimulation}
    A relation $\Rel$ is a multi-resolution approximate bisimulation relation between $\TS$ and $\AbsTS$ (with resolution $\overline{\Rel}$), denoted by $\TS\simeq^{\overline{\Rel}}_{\Rel}\AbsTS$, if:
    \begin{enumerate}
        \item $\Rel$ is a multi-resolution approximate simulation relation from $\TS$ to $\AbsTS$, i.e., $\TS\preceq_{\Rel}^{\overline{\Rel}}\AbsTS$, and
        \item for all $\AbsState\in\AbsStateSet_0$ there exists $\ConState\in\ConStateSet_0$ with $(\ConState,\AbsState)\in\Rel.$$\defdone$
    \end{enumerate}
\end{definition}
Intuitively, \eqref{eq:target-relation} can be thought of as a specification of the coarseness entailed by approximating the dynamics of $\TS$ by those of $\AbsTS$. For example, suppose that  $\epsilon(\ConState,\AbsState) = \PropCoef\min(||\ConState||,||\AbsState||) + \AdditCoef$ for some $\PropCoef,\AdditCoef\in\reals_{\geq0}$; such a relation allows for more coarseness far from the origin, requiring higher resolution near the origin. 

Note that if $\overline{\Rel}$ is obtained by fixing a uniform $\epsilon\in\reals_{\geq0}$, i.e. $\epsilon(\ConState,\AbsState)=\epsilon$ for all $\ConState$ and $\AbsState$ we recover the definition of $\epsilon$-approximate bisimulation relation given in Definition \ref{def:eps-approx}.

\begin{definition}\label{def:solution-space}
    A transition system $\TS$, a resolution relation $\overline{\Rel}$ and a deterministic abstraction of $\TS$: $\AbsTS$, define an \emph{Approximate Relation Synthesis Problem} (ARSP). Formally, we denote the set of its solutions as,
    \begin{equation}
        \SolSpace_{\TS}(\AbsTS, \overline{\Rel}) \doteq \{\Rel \subseteq \ConStateSet\times\ConStateSet : \TS\simeq_{\Rel}^{\overline{\Rel}}\AbsTS\}.
    \end{equation}
    A transition system $\TS$ and a resolution relation $\overline{\Rel}$ define a \emph{Multi-resolution Abstraction Synthesis Problem} (MRASP), with solution space,
    \begin{equation}
        \SolSpace_{\TS}(\overline{\Rel}) \doteq \{(\AbsTS, \Rel) : \TS\simeq_{\Rel}^{\overline{\Rel}}\AbsTS\}.
    \end{equation}
     We overload the notation for $\epsilon$-approximate bisimulation relations and write $\SolSpace_{\TS}(\AbsTS,\epsilon) \doteq \{\Rel : \TS\simeq_{\Rel}^{\epsilon}\AbsTS\}$ and $\SolSpace_{\TS}(\epsilon) \doteq \{(\AbsTS, \Rel) : \TS\simeq_{\Rel}^{\epsilon}\AbsTS\}$.$\defdone$
\end{definition}

\begin{problemstatement}
    Characterise sufficient conditions on $\TS$ and $\overline{\Rel}$ ensuring that $\SolSpace_{\TS}(\overline{\Rel})$ is nonempty. $\defdone$
\end{problemstatement}

The following statements are direct consequences of Definitions \ref{def:eps-approx} and \ref{def:multi-res-bisimulation}. 

\begin{proposition}\label{prop:solution-space}
If the resolution function $\epsilon(\ConState,\AbsState)$ has global extrema then for all $\AbsTS$:
\begin{itemize}
    \item $\SolSpace_{\TS}(\AbsTS, \overline{\Rel})\supseteq \SolSpace_{\TS}(\AbsTS, \epsilon)$, and $\overline{\Rel}$ for all $\epsilon \leq \min_{(\ConState,\AbsState)\in\overline{\Rel}}\epsilon(\ConState,\AbsState)$,
    \item $\SolSpace_{\TS}(\AbsTS, \overline{\Rel})\subseteq \SolSpace_{\TS}(\AbsTS, \epsilon)$, and $\overline{\Rel}$ for all $\epsilon \geq \max_{(\ConState,\AbsState)\in\overline{\Rel}}\epsilon(\ConState,\AbsState)$.$\qed$
\end{itemize}
    
\end{proposition}

\subsection{Incrementally Globally Asymptotically Stable Systems}

\begin{definition}[\cite{tran2016incremental}]
    A system $\TS$ is $\delta$-GAS if there exists a class $\mathcal{KL}$ function $\beta$ such that for all $k \in \naturals$ and $\ConState_0,\ConState_0'\in\ConStateSet_0$ it holds that $
        |\ConState_k - \ConState_k'| \leq \beta(|\ConState_0 - \ConState_0'|, k)$,
    with $\ConState_0\ConState_1\ldots\ConState_k$ and $\ConState_0'\ConState_1'\ldots\ConState_k'$ (finite) behaviours of $\TS$.$\defdone$
\end{definition}

It is well known that $\delta$-GAS systems defined on a compact subset of $\reals$ admit the construction of an abstraction for which it is possible to define an $\epsilon$-approximate bisimulation relation with the original system. We summarise this result below.
\begin{theorem}[Adapted from \cite{tabuada2009verification}]\label{theo:delta-gas-abs}
 For a $\delta$-GAS system $\TS=(\ConStateSet,\ConStateSet_0,\ConFlow)$ with $\ConStateSet\subset\reals^\DimSys$, for any $\epsilon > 0$ there exists a positive scalar $\eta$ and an $\epsilon$-approximate bisimulation relation between $\TS$ and the deterministic abstraction $\AbsTS=(\AbsStateSet,\AbsStateSet_0,\AbsFlow)$ with
$
     \AbsStateSet = \{\ConState\in\ConStateSet : \ConState[i] = k_i \frac{2}{\sqrt{\DimSys}}\eta, k_i\in\mathbb{Z}, i \in \naturals \}.
$  $\qed$
\end{theorem}

The proof of Theorem \ref{theo:delta-gas-abs} relies on the existence of a $\delta$-GAS Lyapunov function $V:\reals^\DimSys\times\reals^\DimSys\rightarrow\reals$. Finding such a function proving that a system is $\delta$-GAS is not a simple task, and in practice it is often expressed as function of $\ConState - \ConState'$, e.g. $V(\ConState,\ConState') = \sqrt{(\ConState - \ConState')^T P (\ConState - \ConState')}$, see \cite{girard2009approximately}. The relation $\Rel$ used to prove the existence of an $\epsilon$-approximate bisimulation is defined as a level set of the $\delta$-GAS Lyapunov function; consequently, given two abstract states $\AbsState\neq\AbsState'$ the sets $\Rel(\AbsState)$ and $\Rel(\AbsState')$ are identical, modulo a translation. This, however, is not strictly required by Definition \ref{def:eps-approx}: we exploit this fact to show that $\delta$-GAS systems are not the only systems accepting approximate bisimulation relations.

From Proposition \ref{prop:solution-space} we obtain the following corollary.

\begin{corollary}
    For every $\delta$-GAS system $\TS$ and for every resolution specification satisfying the conditions of Propostion \ref{prop:solution-space} it holds that $\SolSpace_{\TS}(\overline{\Rel})\neq\emptyset$. $\qed$
\end{corollary}

\subsection{Non-Incrementally Globally Asymptotically Stable Systems}
While most of the existing literature focuses on approximate relations for systems exhibiting some type of `contractive' dynamics, we shift our focus to a simple non-$\delta$-GAS system.

\begin{example}\label{ex:simple-uniform} \emph{($\epsilon$-approx. Bisimulation.)}
    Consider the system $\ConState_{k+1}=2\ConState_k$ with $\ConStateSet=[1,16]$, and suppose we seek an abstraction admitting an $\epsilon$-approximate bisimulation with $\epsilon=1$. The mesh defined by circle diameters and $\AbsStateSet=\{\ConState\in\ConStateSet : \ConState=0.5^k\ConState',;\ConState'\in{9,11,13,15},;0\leq k\leq3\}$ (Figure~\ref{fig:simple-example}) yields a deterministic abstraction (Def.~\ref{def:det-abs}) whose induced relation meets the specification. For $\epsilon=1$, this abstraction is optimal, containing the minimal number of states, exactly 16.$\defdone$
     \begin{figure}[h]
     \centering
     \includegraphics[width=\linewidth]{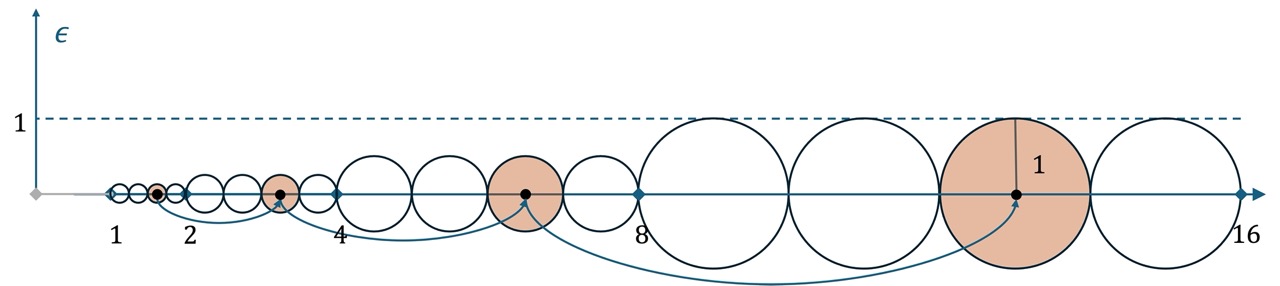}
     \caption{Abstraction satisfying the requested resolution. The abstract states are the midpoints of each segment highlighted by the respective circle, which in turn represent the sought $\epsilon$-approximate relation.}
     \label{fig:simple-example}
 \end{figure}
\end{example}

\begin{figure}[h]
     \centering
     \includegraphics[width=\columnwidth]{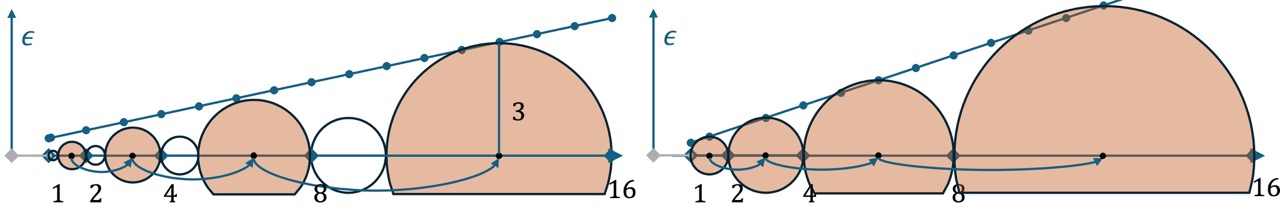}
     \caption{Abstraction satisfying the specified resolution function for Example \ref{ex:simple-non-uniform} (top) and Example \ref{ex:simple-unbounded} (bottom).}\label{fig:simple-example-multi-res}
\end{figure}

\begin{example}\label{ex:simple-non-uniform}\emph{(Multi-res.\ Approx.\ Bisimulation: Bounded Domain.)}
    Consider again the system from Example~\ref{ex:simple-uniform}, but now with design specification $\overline{\Rel}=\{(\ConState,\AbsState)\in\ConStateSet\times\ConStateSet:|\ConState-\AbsState|\leq\epsilon(\ConState,\AbsState)\}$, where $\epsilon(\ConState,\AbsState)=p(\AbsState)=\PropCoef\AbsState+\AdditCoef$, with $p(13)=3$ and $p(1)=0.5$. Constructing a deterministic abstraction with an $\epsilon$-approximate bisimulation would require setting $\epsilon=\min_{\AbsState\in\ConStateSet}p(\AbsState)=0.5$ (Proposition~\ref{prop:solution-space}), yielding an optimal abstraction with 32 states, as shown in Example \ref{ex:simple-uniform}. In contrast, using multi-resolution approximate bisimulation, we can construct an abstraction with only 8 states, satisfying the specification. The solution in Fig.~\ref{fig:simple-example-multi-res} (left) derives from Example~\ref{ex:simple-uniform} by merging states $11$, $13$, and $15$ into a single state at $13$ with $\Rel(13)=[10,16]$, and similarly merging their predecessors. While the abstraction from Example~\ref{ex:simple-uniform} also satisfies the specification, note that $\SolSpace_{\TS}(\overline{\Rel}) \not\supseteq \SolSpace_{\TS}(1)$.$\defdone$
\end{example}

\begin{example}\label{ex:simple-unbounded}\emph{(Multi-res.\ Approx.\ Bisimulation: Unbounded Domain.)}
    Finally, under the same dynamics of Example \ref{ex:simple-uniform}, but with the domain $\ConStateSet = [1, \infty)$, and with resolution specification $\overline{\Rel}$ defined by $\epsilon(\ConState,\AbsState) = \frac{\AbsState}{3}$. A simple solution to the MRASP problem is given by the state set $\AbsState_i = 1.5\cdot2^i$ and $\Rel(\AbsState_i) = [\AbsState_i-2^{i-1}, \AbsState_i+2^{i-1}]$ for $i=0,1,\ldots$, shown in Figure \ref{fig:simple-example-multi-res} (right).
    Since the resolution function grows unbounded over the unbounded domain, Proposition \ref{prop:solution-space} does not apply: indeed, due to the unstable dynamics over the unbounded domain, it is easy to see that $\SolSpace_{\TS}(\epsilon) = \emptyset$ for all $\epsilon>0$, i.e. it is not possible to construct an abstraction that is $\epsilon$-approximately bisimilar to the original system, while considering 
    multi-resolution approximate relations, $\SolSpace_{\TS}(\overline{\Rel})$ admits a solution.$\defdone$
\end{example}

The preceding examples motivate our focus on multi-resolution approximate bisimulation relations. In particular, we aim to exploit the local dynamics and the local resolution to generate compact abstractions, i.e., abstractions satisfying the resolution specification with a minimal number of states. We defer to Section \ref{sec:main-result} the statement and proof of sufficient conditions for general nonlinear systems to admit a solution to an MRASP. We next introduce our second problem formulation.

\section{Compact Abstractions}\label{sec:compact-abstractions}
We frame the search for compact abstractions as a minimisation problem. Typically, the size of an abstraction is measured by the number of transitions. As our focus is on deterministic abstractions, we can equivalently minimise the number of states of the abstraction.

\begin{problemstatement}
Given a TS $\TS$ and a resolution relation $\overline{\Rel}$, find (one of) the smallest deterministic abstractions $\AbsTS=(\AbsStateSet,\AbsStateSet_0,\AbsFlow)$ and relation $\Rel\subseteq\ConStateSet\times\AbsStateSet$ satisfying Definition \ref{def:multi-res-bisimulation}, that is 
$
    \min_{(\AbsTS, \Rel) \in \SolSpace_{\TS}(\overline{\Rel})} |\AbsStateSet|.
$
More explicitly,
\begin{align}\label{eq:min-card-abs}
    \min_{\AbsStateSet \subset \ConStateSet,\Rel}& \quad |\AbsStateSet|  \\
    \text{s.t.} \quad 
    & \forall \ConState \in \ConStateSet_0 \exists\AbsState\in\AbsStateSet. \; (\ConState,\AbsState)\in\Rel, \label{eq:cov}\\
    & (\ConState, \AbsState) \in \Rel \implies (\ConFlow(\ConState), \AbsFlow(\AbsState)) \in \Rel,  \label{eq:con}\\
    & \Rel \subseteq \overline{\Rel}.  \label{eq:res}
\end{align}
Constraints \eqref{eq:cov} to \eqref{eq:res} define the \emph{coverage}, \emph{transition consistency}, and \emph{minimum resolution} conditions, respectively.$\defdone$
\end{problemstatement}

Obtaining a solution to the minimisation problem \eqref{eq:min-card-abs} directly is challenging. 
To address this, we consider a relaxed formulation based on the following assumptions.
\begin{assumption}\label{ass:max-card}
 A maximum budget $\CardAbs$ for the cardinality of $\AbsStateSet\doteq\{\AbsState_1,\ldots,\AbsState_{\CardAbs}\}$ is given. $\defdone$
\end{assumption}
With Assumption~\ref{ass:max-card}, the abstraction size $|\AbsStateSet|$ is fixed in advance, so the minimisation in \eqref{eq:min-card-abs} reduces to checking the feasibility of the constraints.
For tractability, we further restrict the structure of the relation~$\Rel$.
\begin{assumption}\label{ass:rel-template}
    Let $\Rel_{\hat{\boldsymbol{\epsilon}}}\doteq\{(\ConState,\AbsState_i)\in \ConStateSet\times\AbsStateSet: \dist(\ConState,\AbsState_i) \leq \hat{\epsilon}_i\}$ be the template for $\Rel$, parametrised by the $\CardAbs$ positive scalars $\hat{\epsilon}_{i}$'s, one for every $\AbsState_i$, with $\hat{\boldsymbol{\epsilon}}\doteq\hat{\epsilon}_{1},\ldots,\hat{\epsilon}_{k}$. $\defdone$
\end{assumption}
Throughout this work, we restrict attention to templated relations for computational tractability. 
More general templates than the one in Assumption~\ref{ass:rel-template} are considered in Section~\ref{sec:rel-templates-general}. 
Finally, for notational simplicity, we assume $\ConStateSet_0 = \ConStateSet$.

Under Assumptions \ref{ass:max-card} and \ref{ass:rel-template}, the problem can be reformulated into the following relaxed problem. 

\begin{problemrelaxation}\label{problemrelaxation1}
Find $\AbsState_1, \dots, \AbsState_{\CardAbs} \in \ConStateSet$ and $\hat{\epsilon}_1, \dots, \hat{\epsilon}_\CardAbs$ such that, for all $\ConState \in \ConStateSet$, 
\begin{align}
&\Phi_{\text{cov}}(x) \doteq 
   \bigvee_{\mathclap{i = 1}}^\CardAbs
   \left[ d(\ConState, \AbsState_i) \leq \hat{\epsilon}_i \right], 
   \label{eq:coverage} \\[0.2em]
&\Phi_{\text{con}}(x) \doteq 
   \bigwedge_{\mathclap{i=1}}^\CardAbs
   \!\left[ d(\ConState, \AbsState_i) \leq \hat{\epsilon}_i 
   \!\!\implies\!\! d(\ConFlow(\ConState), \AbsState_{g(i)}) \leq \hat{\epsilon}_{g(i)} \right], 
   \label{eq:tran-con} \\[0.2em]
&\Phi_{\text{res}}(x) \doteq 
   \bigwedge_{\mathclap{i=1}}^\CardAbs
   \!\left[ \hat{\epsilon}_i \leq \epsilon(\ConState,\AbsState_i) \right], 
   \label{eq:min-res}
\end{align}
where $g\colon [1,\dots,\CardAbs] \to [1,\dots,\CardAbs]$ maps the index of an abstract state to that of its successor, i.e. $\AbsState_{g(i)} = \AbsFlow(\AbsState_i)$.$\defdone$
\end{problemrelaxation}

The relaxed Problem \ref{problemrelaxation1} represents an MRASP.
In the absence of Assumption \ref{ass:rel-template}, Relaxed Problem \ref{problemrelaxation1} would give rise to a second-order logic formula of the form $\exists \AbsState_1, \dots, \AbsState_{\CardAbs}, \Rel\subseteq\ConStateSet\times\AbsStateSet$ s.t.\ $\forall\ConState \dots$, whose satisfiability is undecidable in general~\cite{DBLP:books/daglib/0082516}.
Even by parametrising the relation $\Rel$, we obtain a formula with a quantifier alternation $\exists \forall $ over a set of conjunctions and disjunctions of (in the best case) linear inequalities, which can be in general  computationally expensive to check even for mature first-order solvers \cite{eirinakis2014quantified}. CEGIS provides a computationally tractable approach by decomposing the quantifier alternation into two subproblems, addressing the existential and universal quantification individually, as detailed in the next section.


\section{Counterexample-Guided Multi-resolution Abstraction Synthesis}\label{sec:cegis-multi-res}

In this section, we describe how we construct a solution for Relaxed Problem \ref{problemrelaxation1}.
We adopt a CEGIS scheme to address the presence of a quantifier alternation. 
Typically, a CEGIS loop involves a \emph{learner} and a \emph{verifier} component. The learner proposes a candidate parameter assignment $\AbsState_1, \dots, \AbsState_{\CardAbs}, \hat{\epsilon}_1, \dots, \hat{\epsilon}_\CardAbs$ satisfying \eqref{eq:coverage}, \eqref{eq:tran-con}, and \eqref{eq:min-res} over a finite number of sample states $\Data \doteq\{x_j\}_{j=1}^N\subseteq\ConStateSet$; this allows to transform the quantifier alternation into a single existential query over a finite set of conjunctions as $\exists \AbsState_1, \dots, \AbsState_{\CardAbs}\in\ConStateSet,
    \hat{\epsilon}_1, \dots, \hat{\epsilon}_\CardAbs \in\reals_{\geq0} \text{ s.t. } $
\begin{equation}\label{eq:single-smt-query}
    \bigwedge_{\ConState_j\in\Data} \Phi_{\text{cov}}(\ConState_j)\wedge\Phi_{\text{con}}(\ConState_j)\wedge\Phi_{\text{res}}(\ConState_j).
\end{equation}
Given the candidate parameter assignment, the verifier checks whether there exists a counterexample in the domain $\ConStateSet$ as $ \exists \ConState\in\ConStateSet $ s.t. $ \neg(\Phi_{\text{cov}}(\ConState)\wedge\Phi_{\text{con}}(\ConState)\wedge\Phi_{\text{res}}(\ConState))$. Observe that, while the learner may use an arbitrary methodology to craft a candidate, potentially based on heuristics, the verifier provides a formal guarantee of correctness in case the check returns no counterexamples. 
The learner, solving \eqref{eq:single-smt-query}, must assign a total of $\CardAbs \cdot n$ real variables for the abstraction's state set and $\CardAbs$ real variables for the scalars $\hat{\epsilon}_i$. While gradient-based methods could be used to craft candidate solutions, encoding the constraints in \eqref{eq:single-smt-query} as a loss function is challenging, particularly due to the implication in the transition consistency condition. 
Alternatively, when $\ConFlow(\ConState)$ and $\dist(\ConState,\AbsState)$ are polynomials,
the learner and the verifier may consist of an SMT solver, which natively handles logical constraints. Unfortunately, obtaining abstractions of practical use typically requires a large $\CardAbs$, making an SMT-based learner excessively slow. Hence, we further split \eqref{eq:single-smt-query} into two simpler problems, where the learner comprises a Clustering and a Relation Learner stage. The verifier consists of a single SMT stage.

\begin{problemrelaxation}\label{problemrelaxation2}
\phantom{.}
\begin{enumerate}
    \item (Clustering) Select $\AbsState_1, \dots, \AbsState_{\CardAbs}\in\ConStateSet$. 
    \item (Relation Learner) Given $\AbsState_1, \dots, \AbsState_{\CardAbs}\in\ConStateSet$, find $\hat{\epsilon}_1, \dots, \hat{\epsilon}_\CardAbs$ s.t. $\forall \ConState \in \ConStateSet$ it holds $\Phi_{\text{cov}}(x) \wedge \Phi_{\text{con}}(x) \wedge \Phi_{\text{res}}(x)$, defined as in \eqref{eq:coverage}, \eqref{eq:tran-con}, \eqref{eq:min-res}.$\defdone$
\end{enumerate}
\end{problemrelaxation}
The second problem relaxation separates the choice of the abstract states $\AbsState_1,\ldots, \AbsState_{\CardAbs}$ from the query yielding the appropriate scalars $\hat{\epsilon}_1,\ldots,\hat{\epsilon}_k$; the former are instead computed beforehand by a fast heuristic, specifically by a clustering algorithm. Problem Relaxation \ref{problemrelaxation2} transforms the MRASP in Relaxed Problem \ref{problemrelaxation1} into 1) finding a deterministic abstraction of $\TS$ with $\CardAbs$ states, driven by heuristics, and 2) solving an ARSP.
This reduces the solution space but, in practice, dramatically speeds up the computation of a solution and enhances the scalability of the approach. Any solution to Relaxed Problem \ref{problemrelaxation2} also constitutes a solution to Relaxed Problem \ref{problemrelaxation1}.

\subsection{Proposed Algorithm - Overview}
In this section, we provide an
overview of the algorithmic scheme illustrated in Figure \ref{fig:cegis-diagram}, designed to solve Relaxed Problem \ref{problemrelaxation2}. From Section \ref{sec:algo-in-detail} onwards, we provide an efficient implementation and details of the scheme.

\subsubsection{Learner}\label{sec:overview-learner}
We begin with an arbitrary finite dataset $\Data \doteq \{\ConState_j\}_{j=1}^N$ of sample states, which serves as input to the learner. 
The learner consists of a clustering stage and an SMT stage. 

\textbf{Clustering Stage.}
The clustering stage uses the dataset to generate $\CardAbs$ anchor points (or abstract states), $\AbsState_1,\ldots\AbsState_\CardAbs$ together with a corresponding partition of $\ConStateSet$ (clusters) $\MeshCell_1,\ldots,\MeshCell_{\CardAbs}$. This yields the mesh $\Mesh\doteq\{(\AbsState_1,\MeshCell_1),\ldots(\AbsState_\CardAbs,\MeshCell_\CardAbs)\}$. Intuitively, each anchor point and corresponding cluster are chosen to group neighbouring data points in $\Data$ sharing similar dynamics. This can be obtained using any clustering algorithm. 
\newline
\emph{Input}: dataset of transitions $\{(\ConState,\ConFlow(\ConState)) : \ConState\in\Data\}$, maximum number of anchor points $\CardAbs$, resolution relation $\hat{\Rel}$. \newline
\emph{Output}: mesh $\Mesh\doteq\{(\AbsState_1,\MeshCell_1),\ldots(\AbsState_\CardAbs,\MeshCell_\CardAbs)\}$.

\textbf{Relation Learner Stage.} The mesh $\Mesh$ uniquely defines the candidate deterministic abstraction $\AbsTS$, according to Definition \ref{def:det-abs}; we are left to find a suitable assignment for the $\hat{\epsilon}_i$'s. Let $Q$ be the equivalence relation $Q\doteq\{(\ConState,\AbsState_i) : \ConState \in \MeshCell_i\}$ induced by the mesh. We exploit $Q$ to conveniently embed the coverage condition $\Phi_{cov}$ by imposing $Q\subseteq \Rel$. One possibility to achieve this is to set $\hat{\epsilon}_i \geq \max_{\ConState\in\MeshCell_i} d(\ConState,\AbsState_i)\doteq\gamma_i$. Then, the query becomes $\exists\hat{\epsilon}_1\geq\gamma_1 \ldots\hat{\epsilon}_\CardAbs\geq\gamma_\CardAbs\in\reals_{\geq0} \text{ s.t. }$
\begin{equation}
     \bigwedge_{\ConState_j\in\Data} \Phi_{\text{con}}(\ConState_j)\wedge\Phi_{\text{res}}(\ConState_j),
\end{equation}
with $\Phi_{\text{con}}(\ConState_j)$ and $\Phi_{\text{res}}(\ConState_j)$ defined as in \eqref{eq:con} and \eqref{eq:res}.
In contrast to Relaxed Problem \ref{problemrelaxation1}, the constraints are much simpler, since the only variables are the $\hat{\epsilon}_i$'s. Under Assumption \ref{ass:rel-template} the scalars $\hat{\epsilon}_1,\ldots,\hat{\epsilon}_k$ uniquely determine the relation. \newline
\emph{Input}: mesh $\Mesh\doteq\{(\AbsState_1,\MeshCell_1),\ldots(\AbsState_\CardAbs,\MeshCell_\CardAbs)\}$, abstraction $\AbsTS$, dataset of transitions $\Data^+=\{(\ConState,\ConFlow(\ConState)) : \ConState\in\Data\}$. \newline
\emph{Output}: candidate relation $\Rel$.

\subsubsection{Verifier}\label{sec:overview-verifier}
The verifier consists of a single SMT query. Given a candidate design of the learner, it checks
\begin{equation}\label{eq:verifier}
    \exists \ConState\in\ConStateSet \ . \ \neg(\Phi_{\text{con}}(\ConState)\wedge\Phi_{\text{res}}(\ConState)).
\end{equation}
Recall that coverage is ensured by a successful assignment of the SMT query in the learner. Additionally, if the resolution function $\epsilon(\ConState,\AbsState)$ depends only on $\AbsState$, e.g. $\epsilon(\ConState,\AbsState) = a|\AbsState| + b$, the verification can be streamlined further: in this case, the check simplifies to $
    \exists \ConState\in\ConStateSet \ . \ \neg\Phi_{\text{con}}(\ConState).
$
\newline
\emph{Input}: candidate abstraction $\AbsTS$, candidate relation $\Rel$. \newline
\emph{Output}: counterexample $\ConState_{ctx}$.

 \begin{figure}[h]
     \centering
     \includegraphics[width=\linewidth]{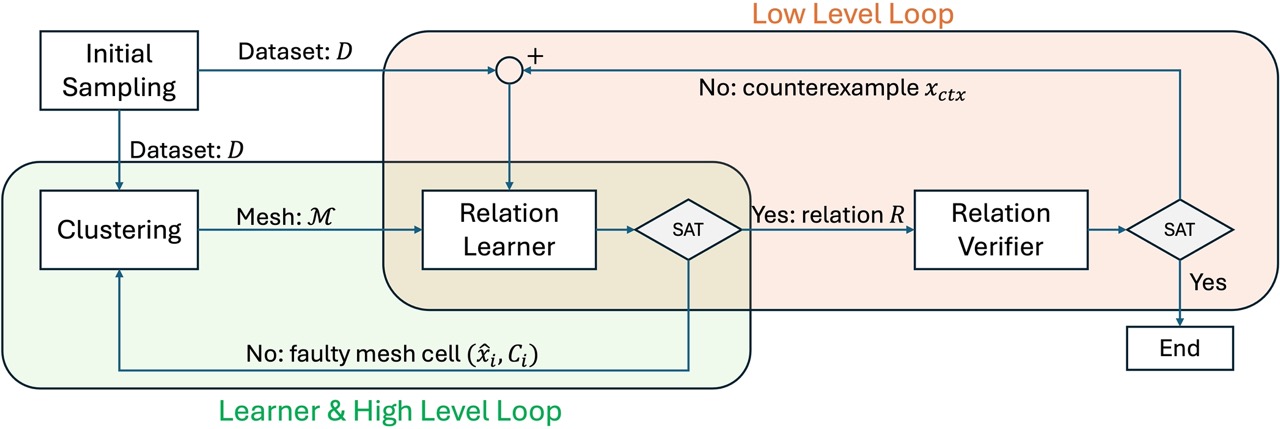}
     \caption{Block diagram of the CEGIS scheme providing a solution to Problem Relaxation \ref{problemrelaxation2}.}
     \label{fig:cegis-diagram} 
 \end{figure}

\subsection{Counterexample Feedback}
If the verifier does not provide a counterexample, the CEGIS loop terminates, having successfully synthesised an abstraction $\AbsTS$ and relation $\Rel$ satisfying Definition \ref{def:multi-res-bisimulation}.
If the verifier provides a counterexample $\ConState_{ctx}$, that is, a point satisfying \eqref{eq:verifier}, it is introduced in the dataset $\Data$; the new dataset $\Data' = \Data\cup\{\ConState_{ctx}\}$ is then fed back to the Relation Learner for a new learning iteration, until the verifier cannot find any more counterexamples. This cycle is highlighted in orange in Figure \ref{fig:cegis-diagram}, and we refer to it as \emph{Low Level Loop}. If the Relation Learner fails at synthesising a candidate relation, the algorithm identifies the set of cells of the current mesh $\Mesh$ and refines it, resulting in a candidate abstraction $\AbsTS$ with an increased number of states, facilitating a successful synthesis as we motivate in the following sections. This cycle is highlighted in green in Figure \ref{fig:cegis-diagram}, and we refer to it as \emph{High Level Loop}. Typically, the High Level Loop iterates a smaller number of times when compared to the Low Level Loop.

\subsection{Proposed Algorithm - In Detail}\label{sec:algo-in-detail}

Referring to Figure \ref{fig:cegis-diagram}, we detail each block in the diagram and its interconnections. While both, the Clustering Stage and the Relation Learner Stage are grouped in the \emph{learner} block, the latter is interlocked directly with the Relation Verifier: these two stages constitute the core of the CEGIS scheme (Low Level Loop). In a nutshell, the Clustering Stage outputs a candidate abstraction $\AbsTS$; next, the Relation Learner designs the candidate relation by addressing the counterexamples of the Relation Verifier, until a multi-resolution $\epsilon$-approximate bisimulation between $\TS$ and $\AbsTS$ is found. The candidate abstraction $\AbsTS$ is updated only when the Relation Learner is unable to find a solution (High Level Loop).

For notational simplicity, we henceforth assume that the resolution function $\epsilon$ depends only on the abstract states, namely $\overline{\Rel} \doteq \{(\ConState,\AbsState) : d(\ConState,\AbsState) \leq \epsilon(\AbsState)\}$.

\begin{figure}[h]
     \centering
         \includegraphics[width=\columnwidth]{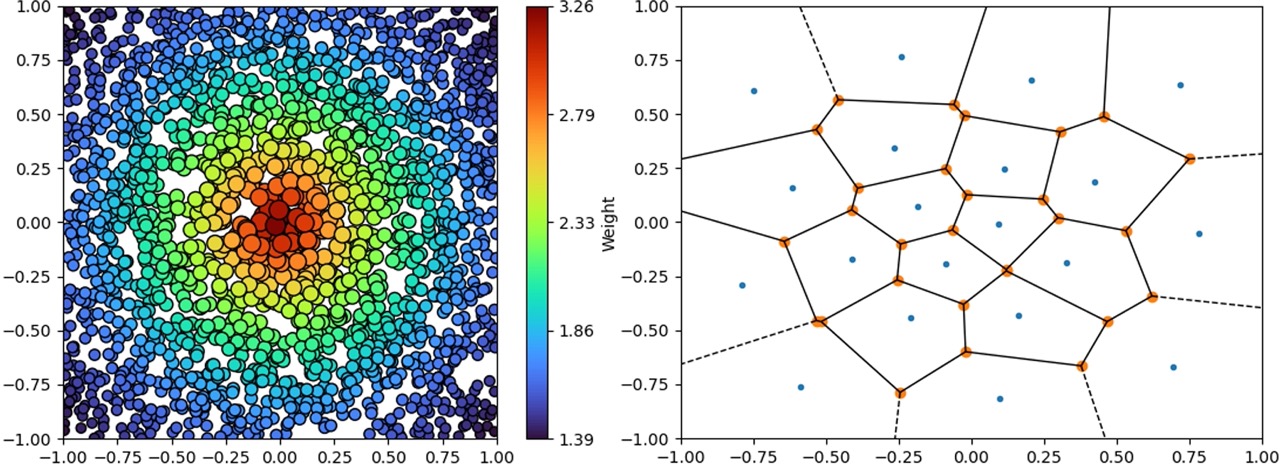}
         \caption{(Left) Effect of the weighting function on the data points: more centroids are drawn near the origin. (Right) Example of a Voronoi diagram resulting from a clustering with 50 centroids.}
         \label{fig:voronoi}
\end{figure}

\textbf{Clustering Stage.} We implement the clustering stage using K-Means~\cite{DBLP:journals/tit/Lloyd82}. We first augment the dataset $\Data$ to $\Data^+\doteq\{(\ConState_j, \ConFlow(\ConState_j))\}_{j=1}^N$. Optionally, we can define a weighting function $w : \ConState \rightarrow \reals$ for the first component of every pair in $\Data^+$, assigning higher weights to data points in locations where higher resolution is requested as specified by the resolution function, resulting in the following clustering cost function
\begin{equation*}
    J(\mu_1,\ldots,\mu_\CardAbs) = \sum_{j=1}^N w(\ConState_j)|| z_j - \mu_{c_j}||,
\end{equation*}
where $z_j = [\ConState_j^T, \ConFlow(\ConState_j)^T]^T$ and $c_j\in [1,...,\CardAbs]$ is the centroid assigned to $z_j$. A simple choice for the weighting function is the inverse of the resolution function, i.e. $w=\epsilon^{-1}$; in Figure \ref{fig:voronoi} (left) we show the effect of such a choice for $\epsilon(\ConState)=0.3|\ConState| + 0.3$ on the domain $[-1,1]^2$. \newline
We define the abstraction's state set using the first $n$ components of each centroid, i.e. $\AbsState_i \doteq \mu_i[1:n]$ for $i\in[1,\ldots,\CardAbs]$, and  
the mesh $\Mesh$ as the Voronoi diagram (under the Euclidean distance) obtained from the abstraction's states
\begin{equation}
    \Mesh \doteq \text{Voronoi}(\AbsState_1,\ldots,\AbsState_k) = \{(\AbsState_1,\MeshCell_1),\ldots(\AbsState_\CardAbs,\MeshCell_\CardAbs)\},
\end{equation}
where each Voronoi cell $\MeshCell_i$ is a convex polytope. As the mesh $\Mesh$ implicitly defines a quantizer $\kappa$, according to Definition \ref{def:det-abs}, the Clustering Stage effectively returns the deterministic abstraction $\AbsTS\doteq(\AbsStateSet,\AbsStateSet,\AbsFlow)$, where $\AbsStateSet\doteq\{\AbsState_1,\ldots,\AbsState_k\}$.

\begin{remark}
    The role of the Clustering Stage is to provide a good first guess for the mesh and the placement of the abstract states. The presence of the High Level Loop ensures that if the Relation Learner can not find a candidate relation, the mesh is suitably refined where needed, as detailed in Section \ref{sec:refinement}. This is critical, since by moving from Relaxed Problem \ref{problemrelaxation1} to Relaxed Problem \ref{problemrelaxation2}, we have separated the selection of abstract states from the selection of the relation. The High Level Loop introduces a means of communication between the two stages; we show that this guarantees the convergence to a solution for a class of systems, see Theorem \ref{theo:non-delta-gas-abs}.$\defdone$
\end{remark}

Given the candidate abstraction $\AbsTS$, it remains to compute the relation $\Rel\subseteq\ConStateSet\times\AbsStateSet$ pairing concrete states $\ConState$ and abstract states $\AbsState$: this is offloaded to the Low Level Loop, as described in the following sections.

\subsection{Relation Templates}\label{sec:rel-templates-general}

From this point onward, we relax Assumption \ref{ass:rel-template}, by allowing more general relation templates for $\Rel$.

\begin{assumption}\label{ass:rel-parameterization}
    Let $\RelCell_{\Params_i}(\AbsState_i)$ denote a compact semialgebraic set in $\ConStateSet$ parametrised by the parameter vector $\Params_i\in\reals^{n_i}$ and associated with $\AbsState_i$, for $i=1,\ldots,k$. We parametrise the space of candidate relations as
    \begin{equation}\label{eq:rel-templatate-general}
    \Rel_{\boldsymbol{\Params},\hat{\boldsymbol{\epsilon}}}\doteq\{(\ConState,\AbsState_i): \ConState\in\RelCell_{\Params_i}(\AbsState_i) \wedge\dist(\ConState,\AbsState_i) \leq \hat{\epsilon}_i\},
    \end{equation}
    where $\boldsymbol{\Params}\doteq \Params_1,\ldots,\Params_k$ and $\hat{\boldsymbol{\epsilon}} \doteq \hat{\epsilon}_1,\ldots,\hat{\epsilon}_k.$ $\defdone$
\end{assumption}

We next discuss different choices for the templates.

\subsubsection{Purely Metric Relation Template}\label{sec:purely-metric-rel}
In the simplest situation, the sets $\RelCell_{\Params_i}(\AbsState_i)$ can themselves be defined by the distance; for instance if $\dist(\ConState,\AbsState_i) = ||\ConState-\AbsState_i||_2$ and $\RelCell_{\Params_i}(\AbsState_i)=||\ConState-\AbsState_i||_2^2 -\Params_i^2\leq0$ (with $\Params_i\in\reals$), then the space of relations reduces to
\begin{equation}\label{eq:rel-simple}
\Rel_{\boldsymbol{\Params},\hat{\boldsymbol{\epsilon}}}\doteq\{(\ConState,\AbsState_i)\in \ConStateSet\times\AbsStateSet:  ||\ConState-\AbsState_i||_2 \leq \min(\hat{\epsilon}_i,\Params_i)\},
    \end{equation}
thus we recover the template in Assumption \ref{ass:rel-template}.

\subsubsection{Polytopic Relation Template}\label{sec:polytopic-rel}

Alternatively, the clustering stage produces a mesh given by the Voronoi diagram induced by the abstract states. Every Voronoi cell admits a half-space representation of the form
\begin{equation*}
    \MeshCell_i = \{\ConState\in\ConStateSet: A_i\ConState \leq b_i, A_i \in \reals^{p\times n}, b_i \in \reals^p\}.
\end{equation*}
The rows of $A_i$ correspond to the normal vectors defining the cell.  
The scaling of a cell $\MeshCell_i$ around a point $\AbsState_i$ by a factor $\Params_i>0$ is defined as
\begin{equation}\label{eq:scaled-cell}
    \RelCell_{\Params_i}(\AbsState_i) \doteq \{\ConState\in\ConStateSet : A_i(\ConState-\AbsState_i) \leq \Params_i (b_i - A_i\AbsState_i)\}.
\end{equation}
Note that for $\Params_i$ equal to 1, $\RelCell_{\Params_i}(\AbsState_i)=\MeshCell_i$; to enforce the coverage condition \eqref{eq:coverage} in the Relation Learner we impose $\Params_i\geq1$, as anticipated in Section \ref{sec:overview-learner}.

\subsection{Low Level Loop - One Shot}\label{sec:low-level-loop-one-shot}
The Low Level Loop, comprising the Relation Learner and the Verifier, encodes and solves by CEGIS the following first-order satisfiability problem
\begin{equation}\label{eq:low-level-loop-problem}
    \exists \Params_1,\ldots,\Params_{\CardAbs}\geq1,\forall\ConState\in\ConStateSet \ . \ \Phi_{\text{con}}(\ConState)\wedge\Phi_{\text{res}}(\ConState).
\end{equation}
The inputs are the initial dataset $\Data$, the candidate abstraction $\AbsTS$, the mesh $\Mesh$, and a relation template $\Rel_{\boldsymbol{\Params},\hat{\boldsymbol{\epsilon}}}$.  Each $\Rel_{\boldsymbol{\Params},\hat{\boldsymbol{\epsilon}}}(\AbsState_i)$ describes a scaled mesh cell as introduced in Section \ref{sec:polytopic-rel}. 
Recall that every $\Params_i$ influences the scaling of the mesh cell $\MeshCell_i$ containing $\AbsState_i$.
By lower bounding each $\Params_i$ by 1, we ensure that the coverage condition always holds. Additionally, let us fix $\hat{\epsilon}_i = \max_{\ConState\in\RelCell_{\Params_i}(\AbsState_i)}\dist(\ConState,\AbsState_i)$; we can rewrite the radius as
\begin{equation}\label{eq:local-resolution-template}
    \hat{\epsilon}_i = \Params_i\max_{\ConState\in\RelCell_{1}(\AbsState_i)}\dist(\ConState,\AbsState_i)=\Params_i\gamma_i.
\end{equation}
We have reduced the number of parameters to be determined in the Low Level Loop to one per cluster. This parallels Relaxed Problem \ref{problemrelaxation2}: instead of directly choosing the radii $\hat{\epsilon}_i$, we equivalently determine the appropriate scaling factors $\Params_i$.

\textbf{Relation Learner.} The Relation Learner aims to solve:
\begin{equation}\label{eq:rel-learner-templated}
    \exists \Params_1,\ldots,\Params_{\CardAbs}\geq 1 \text{ s.t. } \bigwedge_{\ConState_j\in\Data} \Phi_{\text{con}}(\ConState_j)\wedge\Phi_{\text{res}}(\ConState_j),
\end{equation}
where
\begin{align*}
    &\Phi_{\text{con}}(\ConState_j)\doteq\bigwedge_{i=1}^\CardAbs\left[\ConState_j\in \Rel_{\boldsymbol{\Params},\hat{\boldsymbol{\epsilon}}}(\AbsState_i)  \implies \ConFlow(\ConState_j)\in \Rel_{\boldsymbol{\Params},\hat{\boldsymbol{\epsilon}}}(\AbsState_{g(i)})\right],\\
    &\Phi_{\text{res}}(\ConState_j)\doteq\bigwedge_{i=1}^\CardAbs \left[ \Params_i\leq \frac{\epsilon(\AbsState_i)}{\gamma_i} \right].
\end{align*}

For the chosen template, for any $i$, the antecedent $\ConState_j\in \Rel_{\boldsymbol{\Params},\hat{\boldsymbol{\epsilon}}}(\AbsState_i)$ is equivalent to the condition $A_i(\ConState_j - \AbsState_i)\leq\Params_i(b_i-A_i\AbsState_i)$  and the consequent $\ConFlow(\ConState_j)\in \Rel_{\boldsymbol{\Params},\hat{\boldsymbol{\epsilon}}}(\AbsState_{g(i)})$ is equivalent to $A_{g(i)}(f(\ConState_j)-\AbsState_{g(i)})\leq\Params_{g(i)}(b_{g(i)}-A_{g(i)}\AbsState_{g(i)})$.

Observe that, independently of the transition function $\ConFlow$, the predicates in the constraints listed above are linear in $\Params_1,\ldots,\Params_k$. The same applies to any template $\RelCell_{\Params_i}$ that is linear in $\Params_i$. As such, we have encoded the Relation Learner stage in a single existential sentence in the theory of Linear Real Arithmetic ($\exists$-LRA).
This can be efficiently solved with SMT solvers like Z3, see \cite{de2008z3}.

\textbf{Verifier.}
The verifier follows verbatim Section \ref{sec:overview-verifier}. It checks $\exists \ConState\in\ConStateSet \text{ s.t. } \neg\Phi_{\text{con}}(\ConState)$. Where $\Phi_{\text{con}}(\ConState)$ is given by the implication \eqref{eq:con}. Under a polytopic template, the antecedent in the implication above is always linear in the variable $\ConState$; the consequent is linear only when $\ConFlow$ is linear in $\ConState$. If $\ConFlow$ is polynomial, the problem is in the existential fragment of nonlinear real arithmetic ($\exists$-NRA). While this fragment of first-order logic is decidable, Z3's implementation is incomplete, meaning that it is not guaranteed that the search for a counterexample will terminate. For this fragment, Z3 relies on heuristics, and the solution time depends on the number of constraints, number of variables, and degree of polynomials; usually it can handle quickly small (usually less than $6$-$8$ variables), low-degree (usually less than $4$) polynomials. As the number of constraints rapidly grows with the number of abstract states and the dimensionality of the system, in the next Section, we propose an alternative algorithm to alleviate the slowdown due to the number of constraints.

\subsection{Low Level Loop - Parallelised}\label{sec:low-level-loop-parall}

We show that we can find a satisfying assignment to $\boldsymbol{\Params}$ in \eqref{eq:low-level-loop-problem} by solving a sequence of ordered minimisation problems. 
\begin{definition}
    Denote by $\Ancestors_i$ and $\Descendants_i$ the set of abstract states eventually reaching $\AbsState_i$ (ancestors) and the set of abstract states reached by $\AbsState_i$ (descendants) respectively, formally,
    \begin{align}
        \Ancestors_i &\doteq \{\AbsState\in\AbsStateSet : \exists q \in \naturals \ . \ \AbsFlow^q(\AbsState) = \AbsState_i \}, \label{eq:ancestors} \\
        \Descendants_i &\doteq \{\AbsState\in\AbsStateSet : \exists q \in \naturals \ . \ \AbsFlow^q(\AbsState_i) = \AbsState \} \label{eq:descendants}.
    \end{align}
    The graph of the abstraction is a Directed Acyclic Graph (DAG) if for all $\AbsState_i$'s $\Ancestors_i\cap\Descendants_i=\emptyset$. $\defdone$
\end{definition}
With a slight abuse of notation, let $\Rel_{\Params_i}(\AbsState_i) = \Rel_{\boldsymbol{\Params},\hat{\boldsymbol{\epsilon}}}(\AbsState_i)$  to emphasise the fact that the projection of $\Rel_{\boldsymbol{\Params},\hat{\boldsymbol{\epsilon}}}(\AbsState_i)$ at $\AbsState_i$ on $\ConStateSet$ depends solely on the parameter $\Params_i$, and let $g^{-1}:[1,\ldots,k]\rightarrow 2^{[1,\ldots,k]}$ be the inverse of $g$, mapping each index $i$ to the indices of the predecessors of the abstract state $\AbsState_i$. 
\begin{proposition}\label{prop:cascade-of-milps}
    Consider the following subproblem, with minimiser denoted by $\SubProb_i$:
\begin{align}
    \min_{\Params_i\geq1}& \quad \Params_i \label{eq:sub-min-prob}\\
    \text{s.t.} \> 
    & \forall \ConState \in \ConStateSet, \!\!\! \bigwedge_{j\in g^{-1}(i)}\!\!\left[\ConState \in \Rel_{\Params_j^*}(\AbsState_j) \!\!\implies \!\! \ConFlow(\ConState) \in \Rel_{\Params_i}(\AbsState_i)\right]\label{eq:feasibility-sub-prob} \\
    & \Params_i\leq \frac{\epsilon(\AbsState_i)}{\gamma_i}  .\nonumber
\end{align}

If the graph of the abstractions forms a DAG, Problem \eqref{eq:low-level-loop-problem} admits a solution if and only if $\boldsymbol{\Params}^* = \SubProb_1,\ldots,\SubProb_k$ is a solution to \eqref{eq:sub-min-prob} for all $i=1,\ldots,k$. $\qed$
\end{proposition}

Proposition \ref{prop:cascade-of-milps}, proved in Appendix \ref{proofs:prop}, \ref{prop:cascade-of-milps} enables Problem \eqref{eq:low-level-loop-problem} to be solved as a sequence of one-dimensional minimisation problems, exploiting the DAG structure of the abstraction. The solution $\SubProb_i$ depends on $\SubProb_j$ only if there is a path from node j to node i in the DAG. We defer to Section \ref{sec:cyclic-abstractions} a discussion on how Proposition \ref{prop:cascade-of-milps} generalises to cases where the abstraction graph is not a DAG.

\begin{corollary}\label{cor:local-independence}
Let $\Ancestors_i$ be the set of ancestors of $\AbsState_i$ and let $\SubProb_i$ be the solution of \eqref{eq:sub-min-prob}. If $\AbsState_j$ is not an ancestor of $\AbsState_i$ then $\SubProb_i$ does not depend on $\SubProb_j$. $\qed$
\end{corollary}

\begin{remark}
    Corollary \ref{cor:local-independence}, proved in Appendix \ref{proofs:cor}, relies on the imposition of the coverage condition by design, $\Params_i\geq1$. Corollary \ref{cor:local-independence} highlights a computational advantage in that it allows for parallelising the solution of each $\SubProb_i$, according to its dependencies. $\defdone$
\end{remark}

The feasibility set for the first constraint \eqref{eq:feasibility-sub-prob} is given by a logical formula with quantifier alternation ($\exists \Params_i \forall \ConState $). Similarly to Section \ref{sec:low-level-loop-one-shot}, this can be solved by a CEGIS where the Relation Learner is given by a MILP solver solving:
\begin{align}
      \min_{\Params_i\geq1}& \quad \Params_i \label{eq:sub-min-prob-CEGIS}\\
    \text{s.t.} \>
    & \forall \ConState_j \in \Data, \!\!\! \bigwedge_{j\in g^{-1}(i)}\!\!\left[ \ConState_j \in \Rel_{\Params_j^*}(\AbsState_j) \!\!\implies \!\! \ConFlow(\ConState_j) \in \Rel_{\Params_i}(\AbsState_i)\right], \nonumber \\
    & \Params_i\leq \frac{\epsilon(\AbsState_i)}{\gamma_i}.  \nonumber
\end{align}
The verifier checks the candidate solution to the current subproblem. If no counterexamples are found, the algorithm proceeds to the next subproblem. We denote this interplay as '\textproc{CEGISLoop}' in the pseudocode provided in Algorithm \ref{algo:LLL-distributed}. The function '\textproc{TopologicalSort}' applies the usual topological sorting to the abstract states, based on the graph.

\begin{algorithm}[h]
\caption{Low Level Loop - Parallelised}
\label{algo:LLL-distributed}
\begin{algorithmic}[1]
  \Procedure{LLLParallelised}{$\Data^+, \ConFlow, \Mesh, \Rel_{\boldsymbol{\Params}, \hat{\boldsymbol{\epsilon}}}$}
  \State \textbf{Input:} $\Data^+$: transitions; $\ConFlow$: flow; $\Mesh$: mesh; $\Rel_{\boldsymbol{\Params}, \hat{\boldsymbol{\epsilon}}}$: relation template.
  \State \textbf{Output:} $\boldsymbol{\Params}$: parameters; \textbf{None}: refinement needed.
    \State $\AbsTS \gets \Call{BuildAbstraction}{\Mesh,\ConFlow}$
    \State $G \gets \Call{BuildAbstractionGraph}{\AbsTS}$
    \State $ordered\_states \gets \Call{TopologicalSort}{G}$
    \State $\boldsymbol{\Params} \gets \text{empty map}$
    \For{each $\AbsState_i$ in $ordered\_states$}
      \State $\text{dependencies} \gets \Call{GetPredecessors}{G, \AbsState_i}$
      \State $\boldsymbol{\Params}_{deps} \gets \Call{LookupParams}{\boldsymbol{\Params}, \text{dependencies}}$
      \State $\Params_i, \text{success} \gets \Call{CEGISLoop}{(\Data^+, \ConFlow, \Rel_{\boldsymbol{\Params}, \hat{\boldsymbol{\epsilon}}}, \boldsymbol{\Params}_{deps})}$
      \If{not success}
        \State \Return None
      \EndIf
      \State $\boldsymbol{\Params}[\AbsState_i] \gets \Params_i$
    \EndFor
    \State \Return $\boldsymbol{\Params}$
  \EndProcedure
\end{algorithmic}
\end{algorithm}

The following corollary to Proposition \ref{prop:cascade-of-milps} yields a comparison between computing a solution for \eqref{eq:low-level-loop-problem} as illustrated in Section \ref{sec:low-level-loop-one-shot} and as illustrated in this section, by Algorithm \ref{algo:LLL-distributed}.
\begin{corollary}\label{cor:global-min}
    Let $\Params_1,\ldots,\Params_\CardAbs$ be any solution satisfying Problem \ref{eq:low-level-loop-problem} and let $\hat{\epsilon}_1, \ldots, \hat{\epsilon}_\CardAbs$ be the radii according to \eqref{eq:local-resolution-template}; let $\SubProb_1,\ldots,\SubProb_\CardAbs$ be the solution obtained by \eqref{eq:sub-min-prob} and $\hat{\epsilon}_1^*, \ldots, \hat{\epsilon}_\CardAbs^*$ be the corresponding radii. For all $i \leq \CardAbs$
    it holds that $\hat{\epsilon}_i^*\leq\hat{\epsilon}_i$. $\qed$
\end{corollary}
This follows immediately from the proof of Proposition \ref{prop:cascade-of-milps}.
Corollary \ref{cor:global-min} states that, for the given candidate abstractions $\AbsTS$ and relation template $\Rel_{\boldsymbol{\Params}}$, solving Problem \ref{eq:low-level-loop-problem} using Proposition \ref{prop:cascade-of-milps} yields the relation with the highest resolution among the ones in the feasible set.

\subsubsection{General Directed Graphs}\label{sec:cyclic-abstractions}
Consider an abstraction whose graph $G$ does not form a DAG. A set of nodes is called \emph{strongly connected} if each node is reachable from every other node. In deterministic abstractions, a set of strongly connected nodes forms a cycle. In this case, we can compute the \emph{condensation} of $G$, which by construction yields a DAG, whose nodes correspond to the strongly connected components of $G$ grouped into supernodes. 
Let $Y$ denote the indices of a set of $m$ abstract states forming a cycle, and thus grouped into a supernode; then instead of a one-dimensional minimisation problem, we obtain a multi-dimensional problem as 
\begin{align}
\min_{\Params_i\geq1, y\in Y}& \quad \sum_{i\in Y}\Params_i \label{eq:sub-min-prob-with-cycles}\\
    \text{s.t.} \>
    & \forall \ConState_j \in \Data,\! \bigwedge_{i\in Y}\bigwedge_{j\in g^{-1}(i)}\!\!\!\left[ \scalebox{0.87}{$ \ConState_j \in \Rel_{\Params_j^*}(\AbsState_j) 
\!\!\!\implies \!\!\!
\ConFlow(\ConState_j) \in \Rel_{\Params_i}(\AbsState_i)$}\right] \nonumber \\
    & \bigwedge_{i\in Y}\Params_i\leq \frac{\epsilon(\AbsState_i)}{\gamma_i}.  \nonumber
\end{align}
The condensation of $G$ can be used to sort the sequence of minimisation problems, analogously to what was shown in Proposition \ref{prop:cascade-of-milps}. Similarly, it is easy to show that Problem \ref{eq:low-level-loop-problem} admits a solution, if and only if it can be solved by cascading the solution subproblems described by \eqref{eq:sub-min-prob-with-cycles}.

\begin{remark}\label{rem:general-polytopic-template}
    We considered the case where each cell template is parametrised by a single scalar value representing the scaling of the corresponding mesh cell. The discussion natuarlly extends to the case where the template $\RelCell_{\Params_i}(\AbsState_i)$ describes any polytopic template and, accordingly, $\Params_i$ is a vector. $\defdone$
\end{remark}

\subsubsection{Mesh Refinement}\label{sec:refinement}
Leveraging Proposition \ref{prop:cascade-of-milps} to solve Problem \ref{eq:low-level-loop-problem} provides an immediate scheme to refine the abstraction whenever Problem \ref{eq:low-level-loop-problem} is infeasible. If the $i$-th subproblem is infeasible, we select the set of abstract states given by $\AbsState_i$ and all of its ancestors; then we split each corresponding mesh cell $\MeshCell_i$ in two, by adding a new hyperplane perpendicular to the largest dimension of the polytopic cell and containing the old centroid, subsequently placing two new centroids in each sub-region, and update the abstraction $\AbsTS$.

\begin{remark}\label{rem:warm-start-learner}
    Given a data set $\Data$ and the corresponding transitions, it is possible to quickly assess whether the current candidate abstraction $\AbsTS$ \emph{can} satisfy the conditions described in Relaxed Problem \ref{problemrelaxation2}: indeed, we can run the Relation Learner stage, without running the Relation Verifier. The returned parameters $\Params_1',\ldots,\Params_{\CardAbs}'$ will only be valid on the given data set, but they provide a lower bound on the actual values of $\SubProb_1,\ldots,\SubProb_{\CardAbs}$. If any of these lower bounds returns $\hat{\epsilon}_i=\Params_i'\gamma_i$ greater than the locally allowed resolution $\epsilon(\AbsState_i)$, we can already conclude that the candidate abstraction can not satisfy the conditions of Relaxed Problem \ref{problemrelaxation2}. Accordingly, we can either refine the current abstraction as described in Section \ref{sec:refinement}, or increase the budget of abstract states $\CardAbs$ altogether. This step allows to reduce the number of iterations between the High Level Loop and the Low Level Loop, reducing the number of queries to the SMT verifier, and speeding up the construction of a solution. $\defdone$
\end{remark}

\section{Incrementally Uniformly Bounded Systems}\label{sec:main-result}
In this section, we present a class of systems that admit a deterministic abstraction and a multi-resolution bisimulation relation for any resolution relation $\overline{\Rel}$. We rely on the notion of Incrementally Uniformly Bounded systems $(\delta$-UB).
\begin{definition}
    A TS $\TS = (\ConStateSet,\ConStateSet_0, \ConFlow)$ is $\delta$-UB if 
    \begin{equation}
    \forall\ConState_0,\ConState_0'\in\ConStateSet_0,k\in\naturals  \quad \dist(\ConState_k,\ConState_k') \leq \alpha(\dist(\ConState_0,\ConState_0')),\label{eq:bounded-growth}
    \end{equation}
    where $\alpha$ is a class $\mathcal{K}$ function. $\defdone$
\end{definition}
The condition above is an adaptation of \cite[Definition 17]{tran2016incremental} for autonomous systems. Note that $\delta$-UB implies continuity of $\ConFlow$.

\begin{theorem}\label{theo:non-delta-gas-abs}
    Let $\TS = (\ConStateSet,\ConStateSet_0, \ConFlow)$ be a $\delta$-UB TS with an asymptotically stable equilibrium at $\ConState^*$, where $\ConStateSet$ is a compact subset of its region of attraction. 
    Then, for any resolution function $\epsilon(\ConState,\AbsState)$, there exists a finite-state deterministic abstraction $\AbsTS$ and a multi-resolution approximate bisimulation relation between $\TS$ and $\AbsTS$ satisfying $\overline{\Rel}$, i.e., $\SolSpace_{\TS}(\overline{\Rel})\neq\emptyset$. $\qed$
\end{theorem}

The proof of Theorem \ref{theo:non-delta-gas-abs} relies on the existence of an abstraction $\AbsTS$ whose graph has a single cycle at the equilibrium. Then, by providing a refinement scheme for the cells of the underlying mesh that monotonically shrinks the size of the cells, hindering the solution of any Subproblem \ref{eq:sub-min-prob}, we show that by exploiting $\delta$-UB, any desired resolution can be achieved anywhere in the domain, see Appendix \ref{proofs:theo}.

\begin{corollary}\label{cor:transient-sys}
    Let $\TS = (\ConStateSet,\ConStateSet_0, \ConFlow)$ be a $\delta$-UB TS, where $\ConStateSet$ is a compact transient set. Then $\SolSpace_{\TS}(\overline{\Rel})\neq\emptyset$. $\qed$
\end{corollary}
Corollary \ref{cor:transient-sys} drops the requirement for the presence of an asymptotically stable equilibrium, requiring instead that the domain of abstraction $\ConStateSet$ is a transient set: if the system is $\delta$-UB it guarantees that for any specified resolution function, the elicited MRASP always admits a solution. The proof of this result is immediate from the proof of Theorem \ref{theo:non-delta-gas-abs}.
\section{Numerical Examples}\label{sec:num-examples}
\subsection{Linear System}\label{ex:linear-sys}
\begin{figure}[h]
     \centering
     \includegraphics[width=\columnwidth]{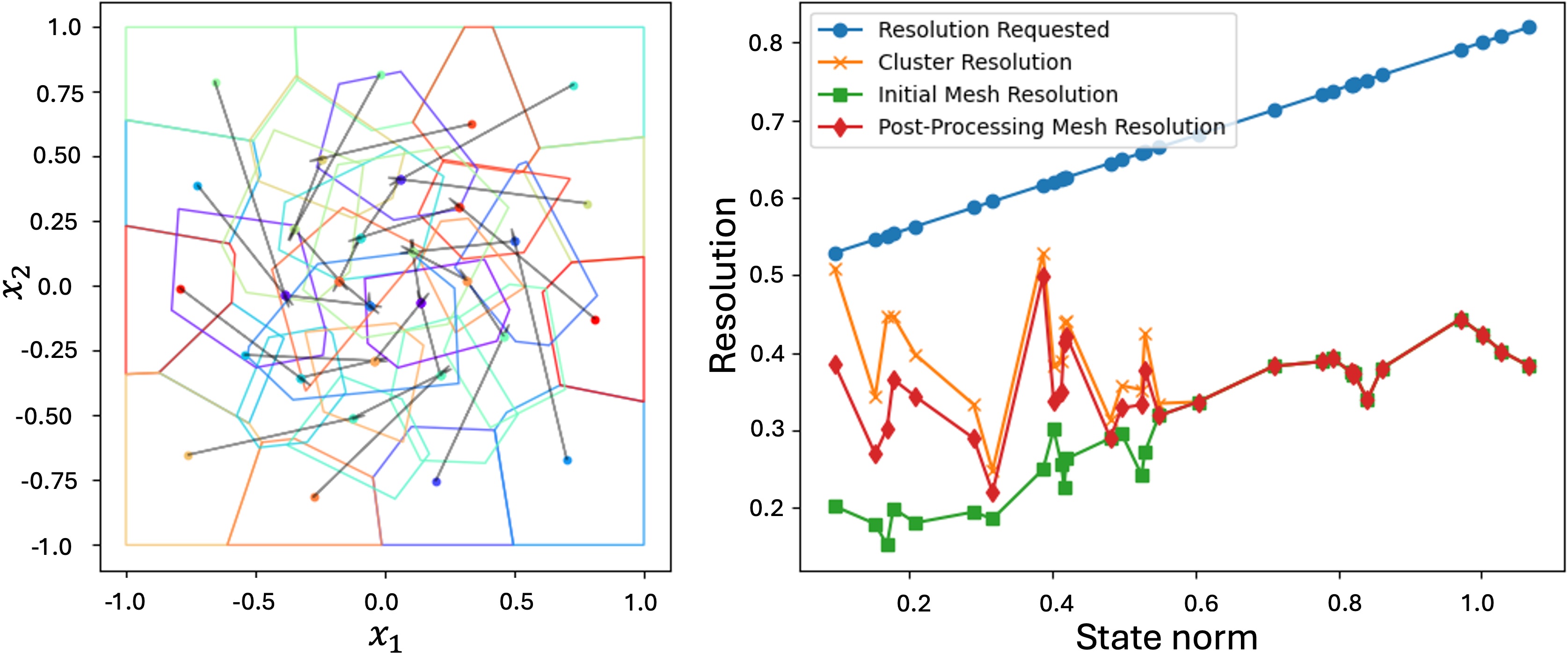}
     \caption{Final relation verified (left), and achieved resolution as a function of the abstract state's norm (right).}\label{fig:resolution-ex1}
\end{figure}

\begin{figure}[h]
     \centering
     \includegraphics[width=\columnwidth]{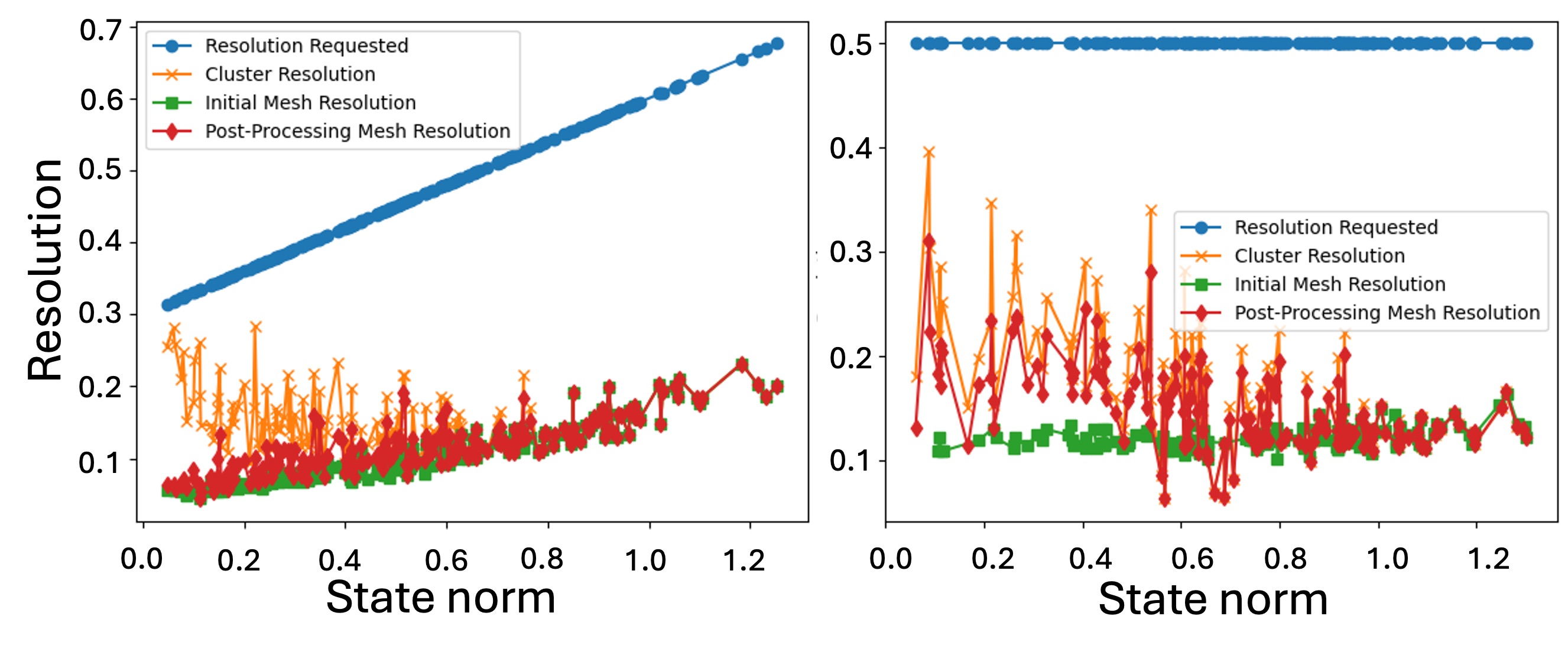}
     \caption{Obtained resolution: Section \ref{ex:linear-sys} with $\CardAbs=200$ (left) and Section \ref{ex:tab-comp}.}\label{fig:resolution-examples}
\end{figure}

Consider the following linear dynamical system
\begin{equation}
    \ConState_{k+1} = 
    0.4\cdot\begin{bmatrix}
    1 & -1 \\
    1 & 1
    \end{bmatrix}\ConState_k.
\end{equation}
The dynamics of the system represents a rotation by $45^{\circ}$ counter-clockwise and scaling by $0.4*\sqrt{2}$ on the domain $\ConStateSet=[-1,1]^2$.
We consider the resolution specification $\overline{\Rel} \doteq \{(\ConState,\AbsState) : ||\ConState-\AbsState||_2 \leq 0.3||\AbsState|| + 0.5\}$. We initialise with $\CardAbs=30$ abstract states and a dataset of $N = 5000$ transitions. In a single iteration, our algorithm constructs an abstraction and a relation $\Rel$ establishing a multi-resolution approximate bisimulation with the concrete system. The synthesis took $\sim 6$ seconds to complete. Figure \ref{fig:resolution-ex1} (left) shows the obtained relation, where each polytope represents the projection of $\Rel$ at a corresponding abstract state. Figure \ref{fig:resolution-ex1} (right) shows how the resolution is obtained as a function of the norm of the abstract states. The green line shows the minimal resolution necessary to guarantee the coverage condition, that is, the resolution of the mesh. The red line shows the lower bound obtained using the pre-processing step described in Remark \ref{rem:warm-start-learner}, and the orange line represents the resolution obtained after the Low Level Loop terminates, which is clearly below the prescribed resolution, indicated by the blue line.
We repeat the experiment using $\CardAbs=200$ and a tighter specification $\overline{\Rel} \doteq \{(\ConState,\AbsState) : ||\ConState-\AbsState||_2 \leq 0.3||\AbsState|| + 0.3\}$. The results are shown in Figure \ref{fig:resolution-examples} (left). The synthesis took $\approx 18$ seconds.
\subsection{Compactness Comparison}\label{ex:tab-comp}
We compare the size of the abstraction resulting from our method with the example in \cite[Example 10.9]{tabuada2009verification}, where the authors consider the system 
\begin{equation}
    \begin{bmatrix}
        \dot{x}\\
        \dot{y}
    \end{bmatrix} = \begin{bmatrix}
        - 7 & 1 \\
        8 & -10
    \end{bmatrix}
    \begin{bmatrix}
        x\\
        y
    \end{bmatrix},
\end{equation}
discretised in time with $\tau = 0.05$. We restrict our focus to the domain $\ConStateSet=[-1,1]^2$. After identifying a $\delta$-GAS Lyapunov function, the authors select a (uniform) resolution of $\epsilon=0.5$, resulting in  $\AbsStateSet = \{\ConState\in\ConStateSet : \ConState[i] = k_i \frac{2}{\sqrt{n}}\eta, k_i\in\mathbb{Z}, i \in \naturals \}$ with $\eta = \frac{\sqrt{2}}{20}$, for a total of 400 abstract states. 
\newline
In contrast, we select as initial guess $k=150$ abstract states, which, after refinement, lead to $171$ abstract states, and a successful synthesis in $\approx17$ seconds. The results are shown in Figure \ref{fig:resolution-examples} (right). Using less than $50\%$ of the abstract states, the obtained relation provides a worst-case resolution $20\%$ higher than the desired value, whereas the rest of the states show an even greater improvement.

\subsection{Non $\delta$-GAS System}\label{ex:non-delta-gas}
Consider the following dynamical system
\begin{equation}
    \begin{bmatrix}
        x_{k+1}\\
        y_{k+1}
    \end{bmatrix} = 
    \begin{bmatrix}
    0.5x_k \\
    0.5y_k + 0.5x_k^2
    \end{bmatrix},
\end{equation}
defined on the domain $\ConStateSet=[-0.5,3]\times[-0.5,5]$. The system is asymptotically stable but not $\delta$-GAS, as can be seen by considering two initial conditions $(x_0,y_0)$ and $(x_0',y_0') = (x_0+\alpha, y_0)$ with $x_0 > 2$. We consider the resolution specification $\overline{\Rel} \doteq \{(\ConState,\AbsState) : ||\ConState-\AbsState||_2 \leq 0.3||\AbsState|| + 0.5\}$. With $k=400$ initial abstract states and $N=10000$ samples, our algorithm successfully synthesises an abstraction and a suitable relation, without requiring any abstraction refinement in 118 seconds.

\begin{figure}[h]
     \centering
     \includegraphics[width=\columnwidth]{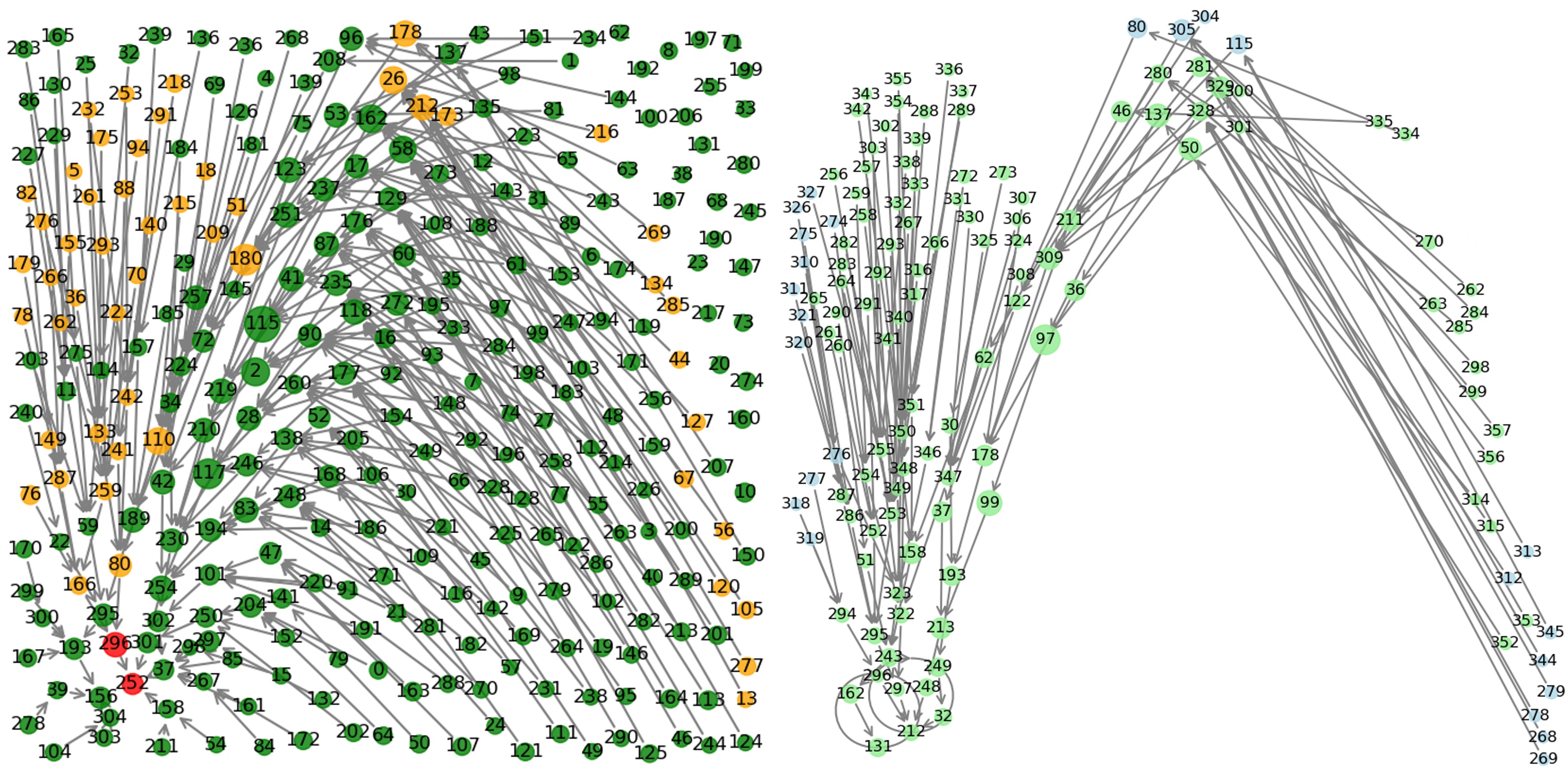}
\caption{Left: failed nodes in Low-Level Loop are highlighted in red, and their ancestors with feasible solutions are depicted in orange. 
Right: refinement of the red and orange nodes.}
     \label{fig:fail-lll}
\end{figure}

\begin{figure}[h]
     \centering
     \includegraphics[width=\columnwidth]{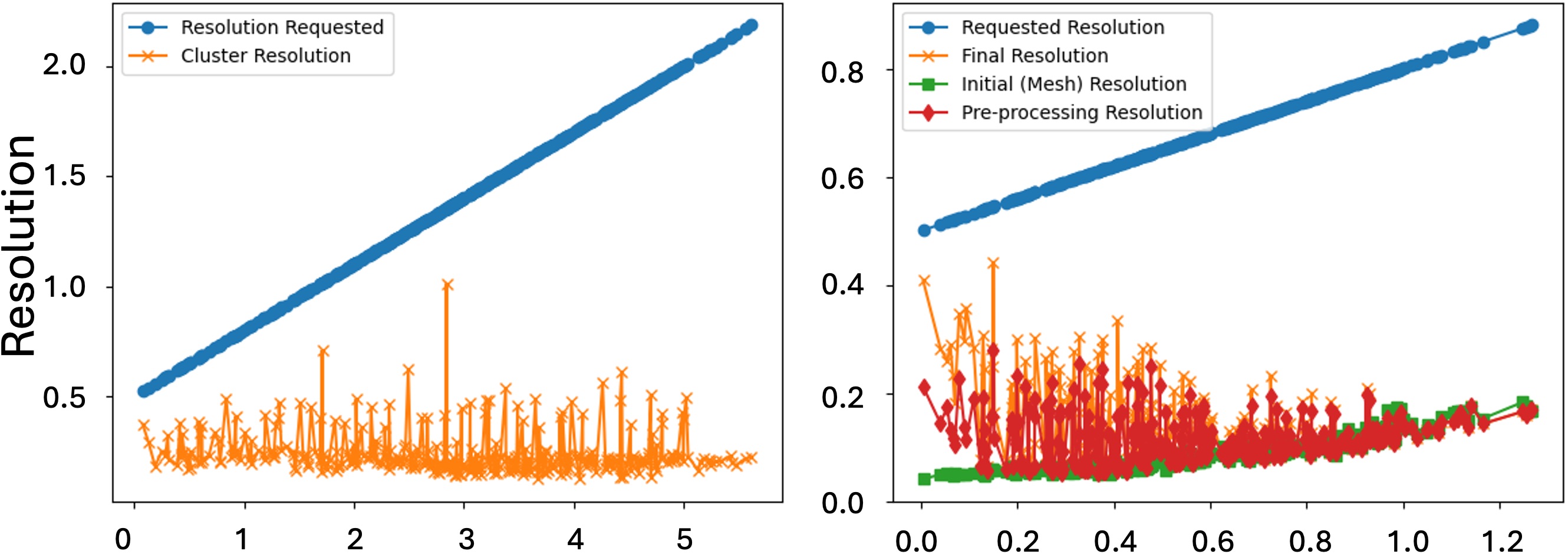}
     \caption{Resolution after refinement of Example \ref{ex:non-delta-gas} (left) and Example \ref{ex:non-diff} (right)}
     \label{fig:resol-after-refinement}
\end{figure}

To illustrate the efficacy of the abstraction refinement protocol, we decrease the initial abstract states to $k=300$. During the first run, the Low Level Loop fails for the abstract state labeled `296' and marked in color red in Figure \ref{fig:fail-lll} (left), which in turn causes the failure of all its descendants, in this case `252'. This triggers an abstraction refinement step: all of the ancestors of node `296' marked in orange undergo a splitting, after which the abstraction's graph is reconstructed. Note that only the set of nodes that were split and all of its descendants need to be resubmitted to the Low Level Loop, shown in Figure \ref{fig:fail-lll} (right). The remaining nodes where already solved in the first run. This greatly reduces the overall solution time. The results are shown in Figure \ref{fig:resol-after-refinement}. The first run took 69 seconds, while the second one only 21 seconds. The final abstraction has 358 abstract nodes.

\subsection{Non Differentiable System}\label{ex:non-diff}
Consider the following dynamical system
\begin{equation}
    \begin{bmatrix}
        x_{k+1}\\
        y_{k+1}
    \end{bmatrix} = 
    \begin{bmatrix}
    0.8x_k \\
    3.2y_k^3
    \end{bmatrix} \text{ if } |y_k| < 0.5,  \text{ else } \begin{bmatrix}
    0.8x_k \\
    0.8y_k
    \end{bmatrix}
\end{equation}
The system is continuous yet nondifferentiable. We choose the resolution specification $\overline{\Rel} \doteq \{(\ConState,\AbsState) : ||\ConState-\AbsState||_2 \leq 0.3||\AbsState|| \! + \! 0.5\}$, and $\CardAbs=300$. In 56 seconds, we obtain a valid abstraction and corresponding relation, as depicted in Figure \ref{fig:resol-after-refinement}. Note that this system cannot be abstracted using the approach of \cite{tazaki2011discrete}.



\subsection{Comparison with \cite{tazaki2010approximately}}
We compare the efficacy of our approach with the algorithm proposed by \cite{tazaki2010approximately}, using the system from the linear system in Section \ref{ex:linear-sys}. We select different resolution specification I) $\epsilon(\ConState)=0.3$, II)  $\epsilon(\ConState)=0.2$, and III)  $\epsilon(\ConState)=0.1$. The results are summarised in Table \ref{tab:comparison}. We observe that the approach of \cite{tazaki2010approximately} provides smaller abstractions compared to our approach. In turn, as the number of abstract states grows, our approach returns a solution in a shorter time.

\begin{table}[h]
\centering
\resizebox{\columnwidth}{!}{%
\begin{tabular}{|c|c|c|c|c|}
\hline
Resolution $\epsilon(\ConState)$ & \#States (Ours) & Time (s, Ours) & \#States (Tazaki et al.) & Time (s, Tazaki et al.) \\
\hline
0.3 & 152  & 9.6   & 60   & 1.5   \\
0.2 & 203  & 8.6   & 124  & 10.0  \\
0.1 & 1431 & 93.4  & 491  & 148.2 \\
\hline
\end{tabular}
}
\caption{Comparison of our approach with \cite{tazaki2010approximately} for different resolution specifications.}
\label{tab:comparison}
\end{table}

\section{Conclusions and Discussion}\label{sec:conclusions}
In this work, we've presented a novel framework for the synthesis of multi-resolution approximate bisimulation relations, designed to construct provably correct abstractions for a broad class of continuous systems. We introduced the first existential proofs for multi-resolution approximate bisimulation relations. By showing that the class of $\delta$-UB systems, which strictly contains $\delta$-GAS systems, admits such relations (Theorem \ref{theo:non-delta-gas-abs}), we extend the class of systems for which approximate bisimulations can be found.

Theorem \ref{theo:non-delta-gas-abs} is constructive, forming the basis of Algorithm \ref{algo:LLL-distributed}, which synthesizes an abstraction $\AbsTS$ and a multi-resolution relation $\Rel$ for any prescribed resolution. For systems with asymptotically stable equilibria, termination may require forward invariant sets around the equilibrium; if the chosen template cannot capture them, the algorithm remains sound but may not terminate. This limitation could be addressed by more flexible polytopic templates (cf. Remark \ref{rem:general-polytopic-template}) or by allowing nondeterminism in the equilibrium cell, which would guarantee termination—directions we leave to future work.

The resulting abstraction size depends on the heuristics in the clustering and mesh refinement stages, yet our experiments show that they are often significantly smaller than those obtained with Lyapunov-based methods \cite{lazar2010infinity}, while neither requiring a Lyapunov function nor explicit growth bounds. Our algorithm instead adapts relation parameters directly from sampled system behaviours in the form of transitions. When abstraction size is not critical and a $\delta$-GAS Lyapunov function is available, Theorem \ref{theo:delta-gas-abs} provides a lightweight alternative.

In contrast with \cite{tazaki2010approximately, tazaki2011discrete}, our approach does not rely on a local linearisation of the system's dynamics, making our algorithm applicable to a larger set of systems. As mentioned \cite{tazaki2011discrete}, computing such linearisations can be computationally expensive for abstractions with a large state set. In fact, differentiability is not strictly required by Theorem \ref{theo:non-delta-gas-abs}, as shown in Example \ref{ex:non-diff}. Moreover, Algorithm \ref{algo:LLL-distributed} exploits the abstraction's graph structure and allows for a parallelised solution. The algorithm presented in \cite{tazaki2010approximately, tazaki2011discrete} is not parallelisable, as the refinement procedure relies on the solution of an optimisation program involving the entire state set of the abstraction. We compared our approach with \cite{tazaki2010approximately} on a linear system, for which \cite{tazaki2010approximately} does not require any linearisation. We observe that for large abstractions, our algorithm can converge to a solution faster. However, the heuristics-based refinement technique presented in \cite{tazaki2010approximately} performs well in practice, yielding smaller abstract state sets. 

Currently, verification dominates runtime, as it relies on SMT solving and scales poorly with system dimension, while learning remains fast. Computation time is also sensitive to clustering quality, since High Level Loop iterations are expensive. Future work will focus on more efficient refinement schemes and tighter coordination between the High and Low Level Loops, aiming to combine the efficiency of \cite{tazaki2010approximately} with our graph-based parallelism. Finally, we plan to extend the approach to nondeterministic multi-resolution abstractions, broadening applicability to systems with unstable equilibria or topological attractors.


\appendix

\subsection{Proposition \ref{prop:cascade-of-milps}}\label{proofs:prop}
$\impliedby$: trivial.

$\implies$: (Sketch, by induction)
Let $\tilde{\boldsymbol{\Params}}=\tilde{\Params}_1,\ldots,\tilde{\Params}_k$ be a solution of \ref{eq:low-level-loop-problem}. 
Let $\AbsState_i$ be a state with no predecessors (at least one such state exists since the abstraction forms a DAG), $\tilde{\Params}_i$ be the associated parameter, $\AbsState_i$ be its abstract successor state with associated parameter $\tilde{\Params}_j$. Since $\AbsState_i$ has no predecessors, from \ref{eq:low-level-loop-problem} we conclude that the only constraint on $\tilde{\Params}_i$ is the resolution condition, i.e. $1\leq\tilde{\Params}_i\leq\epsilon(\AbsState_i)/\gamma_i$, hence $\SubProb_i=1$. If $\tilde{\boldsymbol{\Params}}=\tilde{\Params}_1,\ldots\tilde{\Params}_i,\ldots,\tilde{\Params}_k$ is a solution to \ref{eq:low-level-loop-problem}, then so is $\tilde{\boldsymbol{\Params}}'=\tilde{\Params}_1,\ldots,\SubProb_i,\ldots\tilde{\Params}_k$. Indeed, since $\ConState\in \Rel_{\tilde{\Params}_i}(\AbsState_i)  \implies f(\ConState)\in \Rel_{\tilde{\Params}_j}(\AbsState_j)$ holds, then $\ConState\in \Rel_{\SubProb_i}(\AbsState_i)  \implies f(\ConState)\in \Rel_{\tilde{\Params}_j}(\AbsState_j)$ also holds, since $\Rel_{\SubProb_i}(\AbsState_i)\subseteq \Rel_{\tilde{\Params}_i}(\AbsState_i)$. Similarly, we can set equal to 1 the parameter associated with any abstract state with no predecessors. Now, the predecessors of $\AbsState_j$ have their associated parameters fixed and computed according to \ref{eq:sub-min-prob}. We can repeat the argument inductively to show that we can substitute $\tilde{\Params}_l$ with $\SubProb_l$ to obtain a new improved solution for \ref{eq:low-level-loop-problem}, by solving a cascade of one-dimensional optimization problems following the topological sorting of the abstract states, eventually obtaining the complete solution $\boldsymbol{\Params}^*$. $\qed$

\subsection{Corollary \ref{cor:local-independence}}\label{proofs:cor}
By definition of \eqref{eq:sub-min-prob}, $\SubProb_i$ depends on $\SubProb_j$ if and only if $j\in g^{-1}(i)$; in turn, for any such $j$, $\SubProb_j$ depends on $\SubProb_l$ if and only if $j\in g^{-1}(j)$, and so on. Accordingly, $\SubProb_i$ can be written as a function of the set of $\SubProb_j$'s such that $\AbsState_j\in\Ancestors_i$. $\qed$

\subsection{Theorem \ref{theo:non-delta-gas-abs}}\label{proofs:theo}
We begin by enunciating two lemmas.
\begin{lemma}\label{lem:passing-cell-mesh}
    Let $\Mesh$ be a mesh such that for every cell $\MeshCell$
    \begin{equation}\label{eq:passing-cell}
        \ConState\in\MeshCell \implies \ConFlow(\ConState)\notin\MeshCell.
    \end{equation}
    Let $\Refine^p$ be an operator defined on a pair $(\AbsState_j,\MeshCell_j) \in \Mesh$ as
    \begin{multline}\label{eq:refinement-operator}
        \Refine^p(\AbsState_j,\MeshCell_j) \doteq \{(\AbsState_j^l,\MeshCell_j^l): \MeshCell_j^l\subseteq\MeshCell_j,\\ \AbsState_j^l \in \MeshCell_j^l,  \max_{\ConState\in\MeshCell_j^l}\dist(\ConState,\AbsState_j^l) \leq p \}
    \end{multline}
    such that the set of $\MeshCell_j^l$'s forms a partition of $\MeshCell_j$. 
    It holds that,
    \begin{equation}\label{eq:passing-cell-preserv}
        \ConState \in \MeshCell_j^l \implies \ConFlow(\ConState)\notin\MeshCell_j.
    \end{equation} $\qed$
\end{lemma}
Intuitively, the lemma above states that, if every cell of a mesh $\Mesh$ is such that any state starting in it leaves it in one step, then any subsequent refinement of the $\Mesh$ will preserve this property. We omit the proof.

\begin{lemma}\label{lem:all-sets-transient}
     Let $\TS=(\ConStateSet,\ConStateSet_0,\ConFlow)$ be a TS with asymptotically stable equilibrium at $\ConState^*$, where $\ConStateSet$ is a compact subset of its region of attraction, and $\ConFlow$ is continuous. Then, for all $r>0$ every compact subset $T\subseteq\ConStateSet\setminus\Ball_{r}(\ConState^*)$ is transient. Moreover, any such $T$ admits a mesh $\Mesh$ satisfying \eqref{eq:passing-cell}.
\end{lemma}

\textbf{Proof:} For every $\ConState\in\ConStateSet$ it holds that $\lim_{k\rightarrow\infty}\ConFlow^k(\ConState) = \ConState^*$. Define $\tau(x)\doteq \min\{k \in \naturals : \ConFlow^k(\ConState) \in \Ball_r(\ConState^*)\}$. By asymptotic stability $\tau(\ConState) < \infty$. The function $\tau: T \rightarrow \naturals$ is upper-semicontinuous. To see this note that, for any $k$ $\{\ConState \in T : \tau(\ConState) < k \}$ is open. Indeed the set $\{\ConState \in T : \exists 1 \leq n \leq k \ . \ \ConFlow^n(\ConState) \in \Ball_r(\ConState^*) \}$ is open, as the composition $\ConFlow^n$ of the continuous function $\ConFlow$ is continuous and $\Ball_r(\ConState^*)$ is open. By the Extreme Value Theorem, we conclude that $\sup_{\ConState\in T}\tau(\ConState) = \tau_{max} < \infty$, hence $T$ is transient. Finally, the mesh $\{(\AbsState_i, \MeshCell_i)\}_{i=1}^{\tau_{max}}$ where $\MeshCell_i = \{\ConState\in T: \tau(\ConState) = i\}$ and $\AbsState_i\in\MeshCell_i$ satisfies \eqref{eq:passing-cell}. $\qed$



We proceed now to prove Theorem \ref{theo:non-delta-gas-abs}.

\textbf{Proof:} By Lemma \ref{lem:all-sets-transient}, let $\Mesh$ be a mesh of $\ConStateSet\setminus\Ball_{\nu}(\ConState^*)$ satisfying \eqref{eq:passing-cell}, let $\Mesh'\doteq\Mesh\cup\{(\ConState^*,\Ball_{\nu}(\ConState^*)\}$ be a mesh of $\ConStateSet$ and let  $\AbsTS$ be the resulting deterministic abstraction. 
Consider an arbitrary abstract state $\AbsState_i$, and the template for $\Rel$ given by $\Rel_{\hat{\boldsymbol{\epsilon}}}$ in Assumption \ref{ass:rel-template}. Since we can incorporate the coverage condition by a suitable lower bound $\gamma_i$ for $\hat{\epsilon}_i$ as $\gamma_i = \max_{\ConState\in\MeshCell_i}\dist(\ConState,\AbsState_i)$, we focus on the transition consistency and minimum resolution condition. The transition consistency condition can be rewritten as
\begin{equation}\label{eq:relation-set-inclusion}
    \Rel_{\epsilon_i}(\AbsState_i) \supseteq \bigcup_{j\in g^{-1}(i)} \text{Post}(\Rel_{\epsilon_j}(\AbsState_j)),
\end{equation}
where $g^{-1}(i)$ is the set of indices of $i$'s predecessors. Let $\ConFlow(\ConState)$ be the successor of a point $\ConState$ related to the abstract predecessor $\AbsState_j$. We have,
\begin{align*}
    \dist(\AbsState_{i}, \!\ConFlow(\ConState)) & \!\leq \!\dist(\AbsState_{i}, \ConFlow(\AbsState_j)) \! + \!\dist(\ConFlow(\AbsState_j), \ConFlow(\ConState)) \!\leq \!\gamma_i \!+ \!\alpha(\dist(\AbsState_j,\ConState)), \\
    \hat{\epsilon}_j &\doteq \max_{\ConState\in\Rel_{\hat{\epsilon}_j}(\AbsState_j)} \dist(\AbsState_j,\ConState),
\end{align*}
where the first inequality is a consequence of the triangular inequality and the second inequality follows by the definition of $\gamma_i$ and \eqref{eq:bounded-growth}. Then, setting 
\begin{equation}
    \hat{\epsilon}_i \geq \gamma_i + \max_{j\in g^{-1}(i)}\alpha(\hat{\epsilon}_j)
\end{equation} ensures that \eqref{eq:relation-set-inclusion} is satisfied. Note that $\hat{\epsilon}_j$ is generally greater than its lower bound $\gamma_j$, and can not necessarily be reduced simply by splitting the cell $j$, as $\hat{\epsilon}_j$ depends in turn on its predecessors, and so on. We show now that $\hat{\epsilon}_i$ can be made arbitrarily small.

Denote by $\Ancestors_i$ and $\Descendants_i$ the set of abstract states eventually reaching $i$ (ancestors) and the set of abstract states reached by $i$ (descendants) respectively, according to \eqref{eq:ancestors} and \eqref{eq:descendants}. There are two cases:
\begin{enumerate}
    \item $i$ is not on a cycle: $\Ancestors_i\cap\Descendants_i=\emptyset$
    \item $i$ is on a cycle: $\Ancestors_i\cap\Descendants_i\neq\emptyset$
\end{enumerate}

\textbf{Case 1}: We prove that with a sufficient refinement of the set of ancestors of $i$ we can achieve an arbitrary resolution at $i$.
Let $\Mesh_{\Ancestors_i}$ denote the sub-mesh associated with the ancestors of $i$, i.e. $\Mesh_{\Ancestors_i}\doteq\{(\AbsState_j,\MeshCell_j) \in \Mesh : j\in\Ancestors_i\}$, and let $\tau$ be the transient time of their union. We construct a refinement $\Mesh_{\Ancestors_i}'$ of $\Mesh_{\Ancestors_i}$ by subdividing every pair $(\AbsState_j,\MeshCell_j)$ with the refinement operator defined in \eqref{eq:refinement-operator} as
\begin{equation*}
    \Mesh_{\Ancestors_i}' \doteq \bigcup_{(\AbsState_j,\MeshCell_j)\in\Mesh_{\Ancestors_i}}\Refine^p(\AbsState_j,\MeshCell_j),
\end{equation*}
and let $\AbsTS'$ denote the abstraction obtained by replacing $\Mesh_{\Ancestors_i}$ with $\Mesh_{\Ancestors_i}'$, and let $\Ancestors_i'$ denote the new set of ancestors of $i$. It holds that $\Ancestors_i'\cap\Descendants_i=\emptyset$. Since the refinement does not change the transient time of a set, the new furthest ancestor of $i$ is at distance $q\leq\tau$. Then, setting
\begin{equation}\label{eq:arbitrary-resolution}
    \hat{\epsilon}_i \geq \gamma_i + \Phi^{q}(p)
\end{equation}
satisfies the transition consistency condition for the state $\AbsState_i$ of $\AbsTS'$, where $\Phi(p) = p + \alpha(p)$ is a class $\mathcal{K}$ function of $p$ and $\Phi^{q}$ is the composition of $\Phi$ $q$ times. Since both $\gamma_i$ and $p$ can be made arbitrarily small, we can achieve an arbitrary resolution at $\AbsState_i$, therefore guaranteeing the satisfaction of  $\hat{\epsilon}_i\leq\epsilon(\AbsState_i)$.

\textbf{Case 2}: The set $\Descendants_i$ is the set of states forming the cycle. By assumption, $\TS$ does not admit periodic trajectories. Therefore, the cycle in the abstraction $\AbsTS$ is a spurious behaviour due to the quantisation of abstract the transition function $\AbsFlow\doteq\Quantizer\circ\ConFlow$ and there exists a refinement of $\Descendants_i$ that eliminates the cycle.
Let $\mathcal{C}$ be the region defined by the union of the cells associated with $\Descendants_i$; there exists $q\in\naturals$ such that $\ConFlow^q(\AbsState_i) \notin \mathcal{C}$. It follows that, by adding to the set of abstract states the points $\ConFlow^j(\AbsState_i)$ for $j\leq q$, the cycle is eliminated, leading back to Case 1. 

It remains to address the cell containing the equilibrium $\Ball_r(\ConState^*)$. Let $\AbsState_0\doteq\ConState^*$; clearly $\AbsState_0$ is its own predecessor, that is $0\in g^{-1}(0)$. As we have argued in \eqref{eq:arbitrary-resolution}, the lower bound elicited by the abstract states $\{\AbsState_j\in\AbsStateSet : j\neq 0\in g^{-1}(0)\}$ on $\hat{\epsilon}_0$ can be made arbitrarily small following the same reasoning. Finally, by asymptotic stability, $f$ admits a Lyapunov function $V:\ConStateSet\rightarrow\reals$. Let $\beta = \min_{\dist(\ConState,\ConState^*)=\epsilon(\AbsState_0)} V(\ConState)$.Then the set $\Omega_\beta = \{\ConState\in\ConStateSet : V(\ConState)\leq\beta \}$ is positively invariant. Setting $\Rel(\AbsState_0) = \Omega_{\beta}$ satisfies the condition. Note that, in contrast with the abstract states in $\Mesh$, the concrete states related to the equilibrium $\ConState^*$, $\Rel(\AbsState_0)$, are not necessarily in the form 
 $\{\ConState\in\ConStateSet : \dist(\ConState,\ConState^*)\leq \hat{\epsilon}_0\}$, since $\Omega_\beta$ is a sub-level set of a Lyapunov function.\hfill $\qed$

\bibliography{library}
\bibliographystyle{ieeetr}

\begin{IEEEbiography}
[{\includegraphics[width=1in,height=1.25in,clip,keepaspectratio]{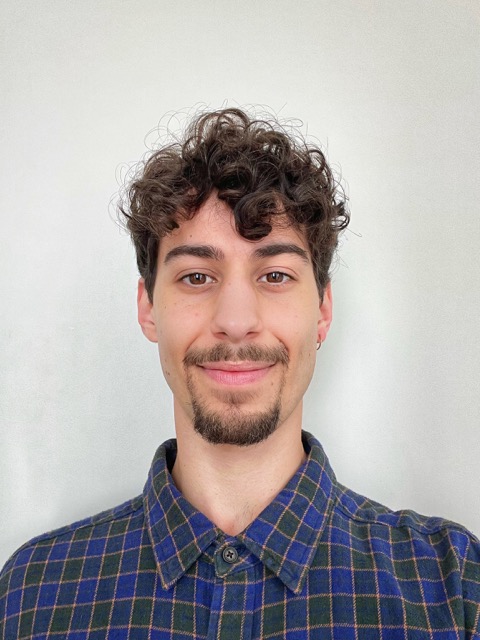}}]{Rudi Coppola} (Student, IEEE) is a PhD candidate at the Delft Center for Systems and Control, in the Delft University of Technology. He obtained a BSc in Electronic Engineering cum laude at the University of Pisa in 2019 and an MSc in Electrical Engineering at King Abdullah University of Science and Technology in 2021.
\end{IEEEbiography}
\begin{IEEEbiography}
[{\includegraphics[width=1in,height=1.25in,clip,keepaspectratio]{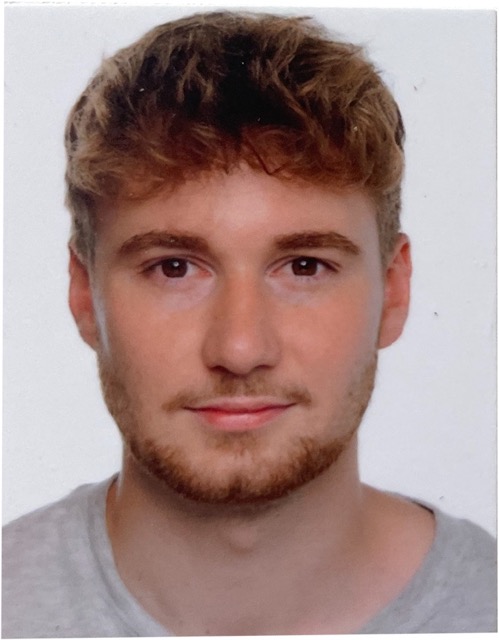}}]{Yannik Schnitzer} is a PhD candidate with the Department of Computer Science, at the University of Oxford, United Kingdom. He obtained a BSc in Computer Science from Saarland University, Saarbr\"ucken, Germany in 2022 and an MSc in Advanced Computer Science from the University of Oxford in 2023.
\end{IEEEbiography}
\begin{IEEEbiography}
[{\includegraphics[width=1in,height=1.25in,clip,keepaspectratio]{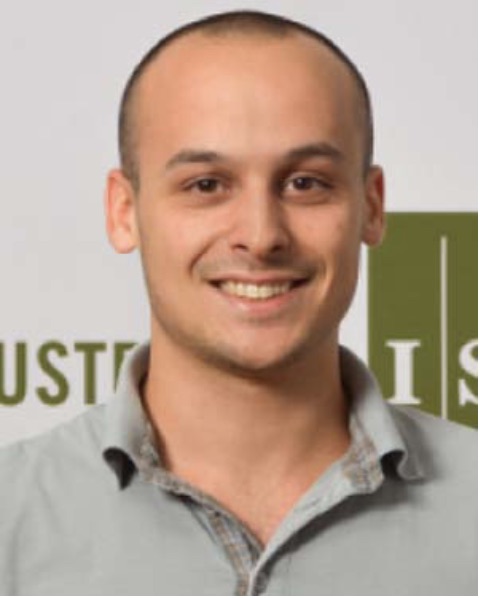}}]{Mirco Giacobbe}  
received the B.S. and M.S. degrees in computer science from the University of
Trento, Trento, Italy, in 2010 and 2012, respectively, the second MSc degree in software systems engineering from RWTH Aachen, Aachen,
Germany, in 2012, and the PhD degree in computer science from IST Austria, Klosterneuburg,
Austria, in 2019.
He is a Lecturer with the School of Computer Science, University of Birmingham, Birmingham, U.K. Before that, he was a Postdoctoral Research Associate with the Department of Computer Science,
University of Oxford, Oxford, U.K. He is interested in formal methods
and machine learning for the analysis and the control of cyber–physical
systems.
\end{IEEEbiography}
\begin{IEEEbiography}
[{\includegraphics[width=1in,height=1.25in,clip,keepaspectratio]{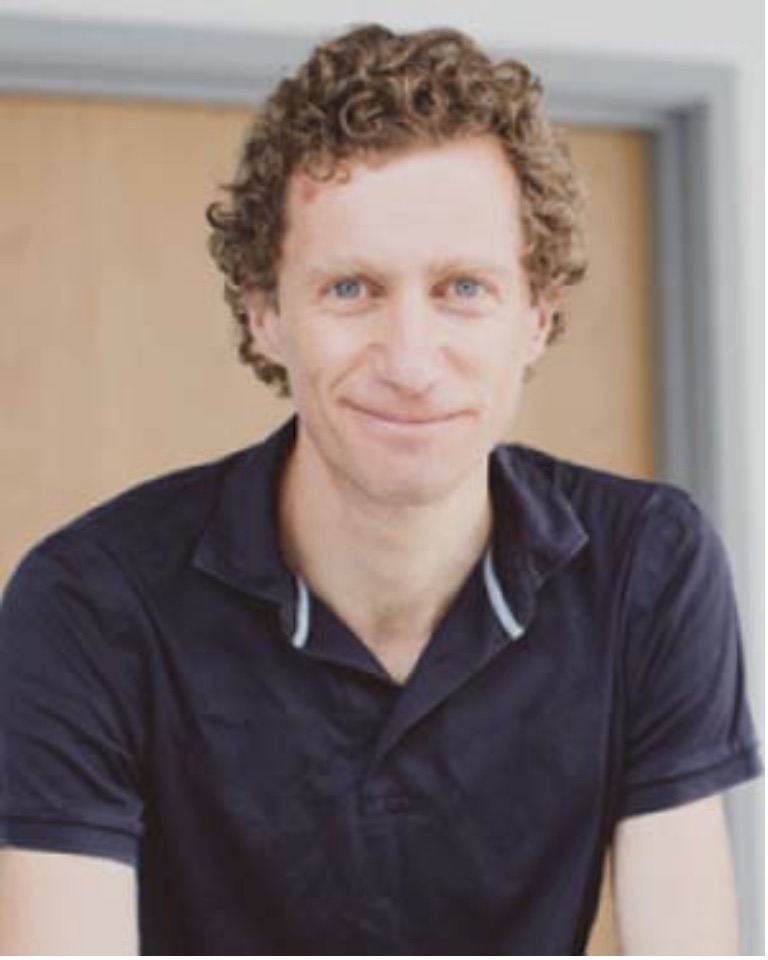}}]{Alessandro Abate} (Fellow, IEEE) received the Ph.D. degree in electrical engineering and computer sciences from the University of California, Berkeley, Berkeley, CA, USA. He is Professor of Verification and Control with the Department of Computer Science, University of Oxford, Oxford, UK.
\end{IEEEbiography}
\begin{IEEEbiography}[{\includegraphics[width=1in,height=1.25in,clip,keepaspectratio]{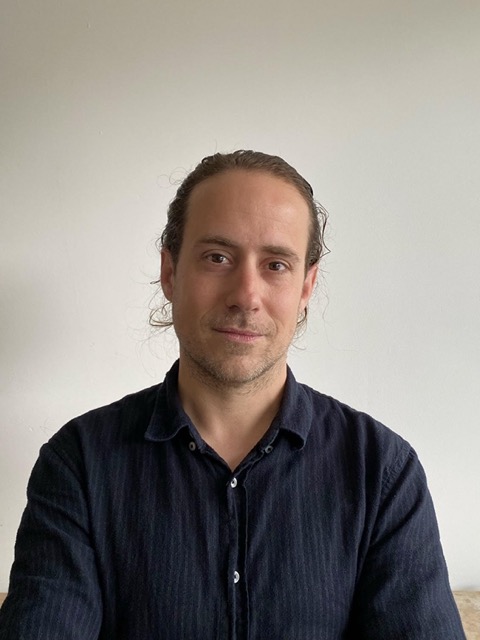}}]{Manuel Mazo Jr.} (Senior Member, IEEE)
is an Associate Professor at the Delft Center for Systems and Control. He obtained an MSc and PhD in Electrical Engineering from the University of California, Los Angeles (UCLA) in 2007 and 2010 respectively. He also holds a Telecommunications Engineering degree from the Polytechnic University of Madrid (UPM) and an Electrical Engineering degree from the Royal Institute of Technology (KTH) in Stockholm, awarded under a TIME double degree program in 2003. Between 2010 and 2012 he held a joint post-doctoral position at the University of Groningen and the innovation centre INCAS3 (The Netherlands). 
He has been the recipient of a University of Newcastle Research Fellowship (2005), the Spanish Ministry of Education/UCLA Fellowship (2005-2009), the Henry Samueli Scholarship from the UCLA School of Engineering and Applied Sciences (2007/2008) and an ERC Starting Grant (2017).
\end{IEEEbiography}

\end{document}